\documentclass[10pt,,draftcls,a4paper,oneside,onecolumn,peerreview]{IEEEtran}
\usepackage[USenglish]{babel} 
\usepackage[T1]{fontenc}
\usepackage[ansinew]{inputenc}
\usepackage{lmodern}

\usepackage[noadjust]{cite}

\usepackage{fancyhdr}
\usepackage{lastpage}
\usepackage{textcomp}
\usepackage{xcolor}

\usepackage{amsmath}
\usepackage{amsthm}
\usepackage{amsfonts}
\usepackage{dsfont} 
\usepackage{stmaryrd} 

\usepackage{graphicx}
\usepackage{multirow}
\usepackage[bookmarks = true, colorlinks = true]{hyperref}

\newtheorem{theorem}{Theorem}
\newtheorem{definition}{Definition}
\newtheorem{conclusion}{Conclusion}
\newtheorem{corollary}{Corollary}
\newtheorem{lemma}{Lemma}
\newtheorem{proposition}{Proposition}

\theoremstyle{definition}
\newtheorem{remark}{Remark}

\DeclareMathOperator*{\argmax}{arg\,max\,}
\DeclareMathOperator*{\argsup}{arg\,sup\,}
\DeclareMathOperator*{\arginf}{arg\,inf\,}
\begin{document}

\title{Two Iterative Proximal-Point Algorithms for the Calculus of Divergence-based Estimators with Application to Mixture Models}
\author{Diaa~Al~Mohamad \; \; \; Michel~Broniatowski,%
\thanks{Diaa Al Mohamad is a PhD. student at Laboratoire de Statistique Th\'eorique et Appliqu\'ee at the Univeristy of Paris 6 (UPMC) 4 place Jussieu 75005 Paris - France. Diaa Al Mohamad is the corresponding author of the article. email: diaa.almohamad@gmail.com}%
\thanks{Michel Broniatowski is a Professor at Laboratoire de Statistique Th\'eorique et Appliqu\'ee at the Univeristy of Paris 6 (UPMC) 4 place Jussieu 75005 Paris - France. Email: michel.broniatowski@upmc.fr}}

\IEEEspecialpapernotice{A part of this work was presented in the conference paper \cite{AlMohamad2015}}

\maketitle

\begin{abstract}
Estimators derived from an EM algorithm are not robust since they are based on the maximization of the likelihood function. We propose a proximal-point algorithm based on the EM algorithm which aim to minimize a divergence criterion. Resulting estimators are generally robust against outliers and misspecification. An EM-type proximal-point algorithm is also introduced in order to produce robust estimators for mixture models. Convergence properties of the two algorithms are treated. We relax an identifiability condition imposed on the proximal term in the literature; a condition which is generally not fulfilled by mixture models. The convergence of the introduced algorithms is discussed on a two-component Weibull mixture and a two-component Gaussian mixture entailing a condition on the initialization of the EM algorithm in order for the later to converge. Simulations on mixture models using different statistical divergences are provided to confirm the validity of our work and the robustness of the resulting estimators against outliers in comparison to the EM algorithm.
\end{abstract}
\begin{IEEEkeywords} 
EM algorithm, mixture model, proximal-point algorithm, robustness, statistical divergence.
\end{IEEEkeywords}

\IEEEpeerreviewmaketitle
\section*{Introduction}
\IEEEPARstart{T}{he} EM algorithm is a well known method for calculating the maximum likelihood estimator of a model where incomplete data are considered. For example, when working with mixture models in the context of clustering, the labels or classes of observations are unknown during the training phase. Several variants of the EM algorithm were proposed, see \cite{McLachlanEM}. Another way to look at the EM algorithm is as a proximal point problem, see \cite{ChretienHero} and \cite{Tseng}. Indeed, one may rewrite the conditional expectation of the complete log-likelihood as a sum of the log-likelihood function and a distance-like function over the conditional densities of the labels provided an observation. Generally, the proximal term has a regularization effect in the sense that a proximal point algorithm is more stable and frequently outperforms classical optimization algorithms, see \cite{Goldstein}. Chr\'etien and Hero \cite{ChretienHeroAccel} prove superlinear convergence of a proximal point algorithm derived by the EM algorithm. Notice that EM-type algorithms usually enjoy no more than linear convergence. Another aspect of proximal point algorithms is that they also permit avoiding saddle points as mentioned in \cite{ChretienHeroProxGener}.\\
Taking into consideration the need for robust estimators, and the fact that the MLE is the least robust estimator among the class of divergence-type estimators which we present below, we generalize the EM algorithm (and the version in \cite{Tseng}) by replacing the log-likelihood function by an estimator of a statistical divergence between the \emph{true distribution} of the data and the model. We are particularly interested in $\varphi-$divergences and the density power divergence which is a Bregman divergence. We recall these two estimators breifly.\\ 
The density power divergence introduced by \cite{BasuMPD} is defined as follows:
\begin{equation}
D_a(g,f) = \int{f^{1+a}(y) - \frac{a+1}{a}g(y)f^a(y)+\frac{1}{a}g^{1+a}(y) dy},\qquad \text{ with } a>0,
\label{eqn:DPD}
\end{equation}
for two probability density functions $f$ and $g$. Given a random sample $Y_1,\cdots,Y_n$ distributed according to some probability measure $P_T$ with density $p_T$ with respect to the Lebesgue measure, and given a model $(p_{\phi})_{\phi\in\Phi}$, the minimum density power divergence estimator (MDPD) is defined by:
\begin{eqnarray}
\hat{\phi}_n & = & \arginf_{\phi\in\Phi} \int{p_{\phi}^{1+a}}(z) dz - \frac{a+1}{a}\frac{1}{n}\sum_{i}^n{p_{\phi}^{a}(Y_i)} \nonumber \\
 & = & \arginf_{\phi\in\Phi} \mathbb{E}_{P_{\phi}}\left[p_{\phi}^a\right] - \frac{a+1}{a}\mathbb{E}_{P_n}\left[p_{\phi}^{a}\right].
\label{eqn:MDPDdef}
\end{eqnarray}
Consistency and robustness properties of the MDPD were studied by \cite{BasuMPD}. The authors show that, the MDPD is generally robust for $a>0$ but the most interesting values of $a$ are in the interval $(0,1)$. Notice that when $a=1$, the MDPD corresponds to the $L^2$ estimator, and as $a$ goes to zero, we obtain the MLE. See \cite{BroniatowskiSeveralApplic} for further properties.\\
A $\varphi-$divergence in the sense of Csisz\'{a}r \cite{Csiszar} is defined (see also \cite{BroniatowskiKeziou2007}) by:
\begin{equation}
D_{\varphi}(Q,P) = \int{\varphi\left(\frac{dQ}{dP}(y)\right)dP(y)},
\label{eqn:phiDivergence}
\end{equation}
where $\varphi$ is a nonnegative strictly convex function and $Q$ and $P$ are two probability measures such that $Q$ is absolutely continuous with respect to $P$. Examples of such divergences are: the Kullback-Leibler (KL) divergence for $\varphi(t)=t\log(t)-t+1$, the modified KL divergence for $\varphi(t)=-\log(t)+t-1$, the hellinger distance for $\varphi(t) = \frac{1}{2}(\sqrt{t}-1)$ among others. All these well-known divergences belong to the class of Cressie-Read functions defined by:
\begin{equation}
\varphi_{\gamma}(t) = \frac{x^{\gamma}-\gamma x + \gamma -1}{\gamma (\gamma -1)}
\label{eqn:CressieReadPhi}
\end{equation}
for $\gamma\in\mathbb{R}\setminus\{0,1\}$ and $\varphi_1(t) = t\log(t)-t+1$ and $\varphi_0(t)=-\log(t)+t-1$. \\
Since the $\varphi-$divergence calculus uses the unknown true distribution, we need to estimate it. We consider the dual estimator of the divergence introduced independently by \cite{BroniaKeziou2006} and \cite{LieseVajdaDivergence}. The use of this estimator is motivated by many reasons. Its minimum coincides with the MLE for $\varphi(t)=-\log(t)+t-1$. Besides, it has the same form for discrete and continuous models, and does not consider any partitioning or smoothing which is not the case of other estimators such as \cite{Beran}, \cite{ParkBasu} and \cite{BasuLindsay} which use kernel density esimators.\\
The dual estimator of the $\varphi-$divergence given an $n-$sample $Y_1,\cdots,Y_n$ is given by:
\begin{equation}
\hat{D}_{\varphi}(p_{\phi},p_T) = \sup_{\alpha\in\Phi}\int{\varphi'\left(\frac{p_{\phi}}{p_{\alpha}}\right)(x)p_{\phi}(x)dx} - \frac{1}{n}\sum_{i=1}^n{\varphi^{\#}\left(\frac{p_{\phi}}{p_{\alpha}}\right)(Y_i)},
\label{eqn:DivergenceDef}
\end{equation}
with $\varphi^{\#}(t)=t\varphi'(t)-\varphi(t)$. Al Mohamad \cite{Diaa} argues that this formula works well under the model, however, when we are not, this quantity largely underestimates the divergence between the true distribution and the model, and proposes following modification:
\begin{equation}
\tilde{D}_{\varphi}(p_{\phi},p_T) = \int{\varphi'\left(\frac{p_{\phi}}{K_{n,w}}\right)(x)p_{\phi}(x)dx} - \frac{1}{n}\sum_{i=1}^n{\varphi^{\#}\left(\frac{p_{\phi}}{K_{n,w}}\right)(Y_i)},
\label{eqn:NewDualForm}
\end{equation}
where $K_{n,w}$ is a nonparametric estimator\footnote{For example, and here in this paper, $K_{n,w}$ is a kernel density estimator based on either symmetric or asymmetric kernel (with or without bias-correction).} of the true distribution $P_T$. The resulting new estimator is robust against outliers. It also permits to get rid of the supremal form which, as we will see later, entails technical and practical issues when one needs to use the continuity or the differentiability of $\hat{D}_{\varphi}(p_{\phi},p_T)$ with respect to $\phi$ in order to prove the convergence of the algorithm.\\
The minimum dual $\varphi-$divergence estimator (MD$\varphi$DE) is defined as the argument of the infimum\footnote{Since there is no guarantee in general that the infimum is unique, the MD$\varphi$DE is defined as any of the points verifying the infimum.} of either $\hat{D}_{\varphi}(p_{\phi},p_T)$ or $\tilde{D}_{\varphi}(p_{\phi},p_T)$. 
\begin{eqnarray}
\text{Classical MD}\varphi\text{DE} & = & \arginf_{\phi\in\Phi} \hat{D}_{\varphi}(p_{\phi},p_T),
\label{eqn:MDphiDEClassique} \\
\text{Kernel-based MD}\varphi\text{DE} & = & \arginf_{\phi\in\Phi} \tilde{D}_{\varphi}(p_{\phi},p_T).
\label{eqn:NewMDphiDE}
\end{eqnarray}
Asymptotic properties and consistency of these two estimators can be found in \cite{BroniatowskiKeziou2007} and \cite{Diaa}. Robustness properties were also studied using the influence function approach in \cite{TomaBronia} and \cite{Diaa}. The kernel-based MD$\varphi$DE (\ref{eqn:NewMDphiDE}) seems to be a \emph{better} estimator than the classical MD$\varphi$DE (\ref{eqn:MDphiDEClassique}) in the sense that the former is robust whereas the later is generally not. Under the model, the estimator given by (\ref{eqn:MDphiDEClassique}) is, however, more efficient.\\
Here in this paper, we propose to calculate the two MD$\varphi$DEs and the MDPD using an iterative procedure based on the work of \cite{Tseng} on the log-likelihood function. This procedure has the form of a proximal point algorithm, and extends the EM algorithm. This algorithm was already introduced and discussed in \cite{AlMohamad2015} and \cite{DiaaBroniaProximalEntropy}. We also propose in this paper a two-step iterative algorithm to calculate the MD$\varphi$DE for mixture models motivated by the EM algorithm. A step to calculate the proportion and a step to calculate the parameters of the components. Proofs for this simplified version become more technical. The goal of this simplification is to reduce the dimension over which we optimize since in lower dimensions, optimization procedures are more efficient\footnote{This does not cover all optimization methods. For example, the Nelder-Mead algorithm is considered as "unreliable" in univariat optimization. The Brent method can be used as an alternative. Note that these two algorithms are suitable for not differentiable functions since they only use function values to reach an optimum.}. Our convergence proof requires some regularity of the estimated divergence with respect to the parameter vector which is not simply checked using (\ref{eqn:DivergenceDef}). Recent results in \cite{Rockafellar} provide sufficient conditions to solve this problem. Differentiability with respect to $\phi$ still remains a very hard task, therefore, our results cover cases when the objective function is not differentiable. 
 \\
Another contribution of this work concerns the assumptions ensuring the convergence of the algorithm. In previous works on such type of proximal algorithms such as \cite{Tseng} and \cite{ChretienHero}, the proximal term is supposed to verify an identifiability property. In other words $D(\phi,\phi')=0$ if and only if $\phi=\phi'$. We show that such property is difficult and it is often not fulfilled in mixture models. We provide a way to relax such assumption in order to cover the case of distance-like functions such as the Kullback-Liebler (the EM case).\\

The paper is organized as follows: We explain in Section \ref{sec:IntrodPart} the context and indicate the mathematical notations which may differ from standard ones. We also present the progression and the derivation of our set of algorithms from the EM algorithm and passing by Tseng's generalization. Section \ref{sec:AnalyticalDiscuss} is devoted to the analytical properties of a supremum function, i.e. a function defined as $\sup_{\alpha}f(\alpha,\phi)$ which is the case of the dual representation of the divergence presented above. In section \ref{sec:Proofs}, we prove some convergence properties of the sequence generated by our algorithm. We show in Section \ref{sec:Examples} and by examples, how one can prove convergence of the proposed algorithms in Gaussian and Weibull mixtures including a convergence proof of the EM algorithm. Finally, Section \ref{sec:Simulations} gives some experimental results confirming the validity of the methods proposed in comparison simply to the maximum likelihood estimator calculated through the EM algorithm.

\section{A description of the algorithm}\label{sec:IntrodPart}
\subsection{General context and notations}
Let $(X,Y)$ be a couple of random variables with joint probability density function $f(x,y|\phi)$ parametrized by a vector of parameters $\phi\in\Phi\subset\mathbb{R}^d$. Let $(X_1,Y_1),\cdots,(X_n,Y_n)$ be n copies of $(X,Y)$ independently and identically distributed. Finally, let $(x_1,y_1),\cdots,(x_n,y_n)$ be n realizations of the n copies of $(X,Y)$. The $x_i$'s are the unobserved data (labels) and the $y_i$'s are the observations. The vector of parameters $\phi$ is unknown and need to be estimated. \\
The observed data $y_i$ are supposed to be real vectors and the labels $x_i$ belong to a space $\mathcal{X}$ not necessarily finite unless mentioned otherwise. Denote $dx$ the measure on the label space $\mathcal{X}$ (for example the counting measure if $\mathcal{X}$ is discrete). The marginal density of the observed data is given by $p_{\phi}(y)=\int{f(x,y|\phi)}dx$.\\
For a parametrized function $f$ with a parameter $a$, we write $f(x|a)$. We use the notation $\phi^k$ for sequences with the index above. Derivatives of a real valued function $\psi$ defined on $\mathbb{R}$ are written as $\psi',\psi'',$ etc. We use $\nabla f$ for the gradient of real function $f$ defined on $\mathbb{R}^d$, $\partial f$ to its subgradient and $J_f$ to the matrix of second order partial derivatives. For a generic function $H$ of two variables $(\phi,\theta)$, $\nabla_1 H(\phi,\theta)$ denotes the gradient with respect to the first (vectorial) variable $\phi$. 
\subsection{EM algorithm and Tseng's generalization}
The EM algorithm is a well-known method for calculating the maximum likelihood estimator of a model where incomplete data are considered. For example, when working with mixture models in the context of clustering, the labels or classes of observations are unknown during the training phase. Several variants of the EM algorithm were proposed, see \cite{McLachlanEM}. The EM algorithm estimates the unknown parameter vector by generating the sequence (see \cite{Dempster}):
\begin{eqnarray*}
\phi^{k+1} & = & \argmax_{\Phi} Q(\phi,\phi^k) \\
					 & = & \argmax_{\Phi} \mathbb{E}\left[\log(f(\textbf{X},\textbf{Y}|\phi)) \left| \textbf{Y}=\textbf{y},\phi^k\right.\right],
\end{eqnarray*}
where $\textbf{X} = (X_1,\cdots,X_n)$, $\textbf{Y} = (Y_1,\cdots,Y_n)$ and $\textbf{y}=(y_1,\cdots,y_n)$. By independence between the couples $(X_i,Y_i)$'s, the previous iteration may be rewritten as:
\begin{eqnarray}
\phi^{k+1} & = & \argmax_{\Phi} \sum_{i=1}^n{\mathbb{E}\left[\log(f(X_i,Y_i|\phi)) \left| Y_i=y_i,\phi^k\right.\right]} \nonumber\\
					 & = & \argmax_{\Phi} \sum_{i=1}^n\int_{\mathcal{X}}{\log(f(x,y_i|\phi)) h_i(x|\phi^k) dx},
\label{eqn:EMAlgo}
\end{eqnarray}
where $h_i(x|\phi^k)$ is the conditional density of the labels (at step $k$) provided $y_i$. It is given by:
\begin{equation}
h_i(x|\phi^k) = \frac{f(x,y_i|\phi^k)}{p_{\phi^k}(y_i)}.
\label{eqn:ConditionalDensLabel}
\end{equation}
This justifies the recurrence equation given by \cite{Tseng}. It is slightly different from the EM recurrence defined in \cite{Dempster}. The conditional expectation of the logarithm of the complete likelihood provided the data and the parameter vector of the previous iteration is calculated, here, on the vector of observed data. The expectation is replaced by an integral against the corresponding conditional density of the labels. \\
It is well-known that the EM iterations can be rewritten as a difference between the log-likelihood and a \emph{Kullback-Liebler} distance-like function. Indeed, using (\ref{eqn:ConditionalDensLabel}) in (\ref{eqn:EMAlgo}), one can write:
\begin{eqnarray*}
\phi^{k+1} & = & \argmax_{\Phi} \sum_{i=1}^n\int_{\mathcal{X}}{\log\left(h_i(x|\phi)\times p_{\phi}(y_i)\right) h_i(x|\phi^k) dx} \\
 					 & = & \argmax_{\Phi} \sum_{i=1}^n\int_{\mathcal{X}}{\log\left(p_{\phi}(y_i)\right) h_i(x|\phi^k) dx} + \sum_{i=1}^n\int_{\mathcal{X}}{\log\left(h_i(x|\phi)\right) h_i(x|\phi^k) dx} \\
           & = & \argmax_{\Phi} \sum_{i=1}^n{\log\left(p_{\phi}(y_i)\right)} + \sum_{i=1}^n\int_{\mathcal{X}}{\log\left(\frac{h_i(x|\phi)}{h_i(x|\phi^k)}\right) h_i(x|\phi^k) dx}\\
					& & \qquad \qquad \qquad \qquad + \sum_{i=1}^n\int_{\mathcal{X}}{\log\left(h_i(x|\phi^k)\right) h_i(x|\phi^k) dx}.
\end{eqnarray*}
The final line is justified by the fact that $h_i(x|\phi)$ is a density, therefore it integrates to 1. The additional term does not depend on $\phi$ and, hence, can be omitted. We now have the following iterative procedure:
\begin{equation}
\phi^{k+1} = \argmax_{\Phi} \sum_{i=1}^n{\log\left(p_{\phi}(y_i)\right)} + \sum_{i=1}^n\int_{\mathcal{X}}{\log\left(\frac{h_i(x|\phi)}{h_i(x|\phi^k)}\right) h_i(x|\phi^k) dx}.
\label{eqn:EMProximal}
\end{equation}
As stated in \cite{Tseng}, the previous iteration has the form of a proximal point maximization of the log-likelihood, i.e. a perturbation of the log-likelihood by a (modified) Kullback distance-like function defined on the conditional densities of the labels. Tseng proposed to generalize the Kullback distance-like term into other types of divergences. Tseng's recurrence is now defined by:
\begin{equation}
\phi^{k+1} = \argsup_{\phi} J(\phi) - D_{\psi}(\phi,\phi^k),
\label{eqn:TsengAlgo}
\end{equation}
where $J$ is the log-likelihood function and $D_{\psi}$ is a distance-like function defined on the conditional probabilities of the classes provided the observations and is given by:
\begin{equation}
D_{\psi}(\phi,\phi^k) = \sum_{i=1}^n\int_{\mathcal{X}}{\psi\left(\frac{h_i(x|\phi)}{h_i(x|\phi^k)}\right)h_i(x|\phi^k)dx},
\label{eqn:DivergenceClassesNtNorm}
\end{equation}
for a real positive convex function $\psi$ such that $\psi(1)=\psi'(1)=0$. $D_{\psi}(\phi_1,\phi_2)$ is positive and equals zero if $\phi_1=\phi_2$. Moreover, $D_{\psi}(\phi_1,\phi_2)=0$ if and only if $\forall i, h_i(x|\phi_1) = h_i(x|\phi_2)$ $dx-$almost everywhere. Clearly, (\ref{eqn:TsengAlgo}) and (\ref{eqn:EMProximal}) are equivalent for $\psi(t)=-\log(t)+t-1$.\\

\subsection{Generalization of Tseng's algorithm}\label{subsec:OurAlgo}
We use the relation between maximizing the log-likelihood and minimizing the Kullback-Liebler divergence to generalize the previous algorithm. We therefore replace the log-likelihood function by a divergence $D$ between the true density of the data $p_T$ and the model $p_{\phi}$. This divergence will either denote density power divergence $D_a$ given by (\ref{eqn:DPD}) or a $\varphi-$divergence given by (\ref{eqn:phiDivergence}). Since the value of the divergence depends on the true density which is unknown, an estimator of the divergence needs to be considered. For the density power divergence, we use the optimized function in (\ref{eqn:MDPDdef}). For $\varphi-$divergences, we use the dual estimator of the divergence defined earlier by either (\ref{eqn:DivergenceDef}) or (\ref{eqn:NewDualForm}). Denote $\hat{D}$ for the estimator of the corresponding divergence. Our new algorithm is defined by the following recurrence:
\begin{equation}
\phi^{k+1} = \arginf_{\phi} \hat{D}(p_{\phi},p_T) + \frac{1}{n}D_{\psi}(\phi,\phi^k)
\label{eqn:DivergenceAlgoPreVersion}
\end{equation}
where $D_{\psi}(\phi,\phi^k)$ is defined by (\ref{eqn:DivergenceClassesNtNorm}). This algorithm was proposed in \cite{DiaaBroniaProximalEntropy} in the context of $\varphi-$divergences. There is how ever no problem in defining the same algorithm for any statistical divergence family which generates the MLE. When $\varphi(t) = -\log(t)+t-1$ or when $a\rightarrow 0$, it is easy to see that we get recurrence (\ref{eqn:TsengAlgo}). Take for example the case of the approximation (\ref{eqn:DivergenceDef}). Since $\varphi'(t) = \frac{-1}{t} + 1$, we have $\int{\varphi'\left(\frac{p_{\phi}}{p_{\alpha}}\right)p_{\phi}dx} = 0$. Hence,
\[\hat{D}_{\varphi}(p_{\phi},p_T) = \sup_{\alpha} \frac{1}{n}\sum_{i=1}^n{\log(p_{\alpha}(y_i))} - \frac{1}{n}\sum_{i=1}^n{\log(p_{\phi}(y_i))}.\]
Using the fact that the first term in $\hat{D}_{\varphi}(p_{\phi},p_T)$ does not depend on $\phi$, so it does not count in the $\arginf$ defining $\phi^{k+1}$, we may rewrite (\ref{eqn:DivergenceAlgoPreVersion}) as:
\begin{eqnarray*}
\phi^{k+1} & = & \arginf_{\phi}\left\{\sup_{\alpha} \frac{1}{n}\sum_{i=1}^n{\log(p_{\alpha}(y_i))} - \frac{1}{n}\sum_{i=1}^n{\log(p_{\phi}(y_i))} + \frac{1}{n}
D_{\psi}(\phi,\phi^k)\right\} \\
           & = & \arginf_{\phi}\left\{-\frac{1}{n}\sum_{i=1}^n{\log(p_{\phi}(y_i))} +\frac{1}{n}D_{\psi}(\phi,\phi^k)\right\} \\
           & = & \argsup_{\phi}\left\{\frac{1}{n}\sum_{i=1}^n{\log(p_{\phi}(y_i))} - \frac{1}{n}D_{\psi}(\phi,\phi^k)\right\} \\
           & = & \argsup_{\phi} J(\phi) - D_{\psi}(\phi,\phi^k).
\end{eqnarray*}
For notational simplicity, from now on, we redefine $D_{\psi}$ with a normalization by $n$, i.e. 
\begin{equation}
D_{\psi}(\phi,\phi^k) = \frac{1}{n} \sum_{i=1}^n\int_{\mathcal{X}}{\psi\left(\frac{h_i(x|\phi)}{h_i(x|\phi^k)}\right)h_i(x|\phi^k)dx}.
\label{eqn:DivergenceClasses}
\end{equation}
Hence, our set of algorithms is redefined by:
\begin{equation}
\phi^{k+1} = \arginf_{\phi} \hat{D}(p_{\phi},p_T) + D_{\psi}(\phi,\phi^k).
\label{eqn:DivergenceAlgo}
\end{equation}
We will see later that this iteration forces the estimated divergence to decrease and that under suitable conditions, it converges to a (local) minimum of $\hat{D}(p_{\phi},p_T)$. It results that, algorithm (\ref{eqn:DivergenceAlgo}) is a way to calculate the divergence-based estimators (\ref{eqn:MDPDdef}), (\ref{eqn:MDphiDEClassique} ) and (\ref{eqn:NewMDphiDE}).\\
Before proceeding to study the convergence properties of such algorithm, we will propose another algorithm for the case of mixture models. In the EM algorithm, the estimation of the parameters of a mixture model is done mainly by two steps. The first step estimates the proportions of the classes whereas the second step estimates the parameters defining the classes. Our idea is based on a directional optimization of the objective function in (\ref{eqn:DivergenceAlgo}). Convergence properties of the two-step algorithm will also be studied, but the proofs are more technical.
\subsection{A two-step EM-type algorithm for mixture models}\label{sec:TwoStepAlgo}
Let $p_{\phi}$ be a mixture model with $s$ components:
\begin{equation}
p_{\phi}(y) = \sum_{i=1}^s{\lambda_i f_i(y|\theta_i)}.
\label{eqn:MixModelDef}
\end{equation}
Here, $\phi = (\lambda,\theta)$ with $\lambda = (\lambda_1,\cdots,\lambda_s)\in [0,1]^s$ such that $\sum_j{\lambda_j}=1$, and $\theta = (\theta_1,\cdots,\theta_s)\in\Theta\subset\mathbb{R}^{d-s}$ such that $\Phi \subset [0,1]^s\times\Theta$. In the EM algorithm, the corresponding optimization to (\ref{eqn:DivergenceAlgo}) can be solved by calculating an estimate of the $\lambda$'s as the proportions of classes, and then proceed to optimize on the $\theta$'s (see for example \cite{Titterington}). This simplifies the optimization in terms of complexity (optimization in lower spaces) and clarity (separate proportions from classes parameters). We want to build an algorithm with the same property and divide the optimization problem into two parts. One which estimates the proportions $\lambda$ and another which estimates the parameters defining the form of each component $\theta$. We propose the following algorithm:
\begin{eqnarray}
\lambda^{k+1} & = & \arginf_{\lambda\in[0,1]^s, s.t. (\lambda,\theta^k)\in\Phi} \hat{D}(p_{\lambda,\theta^k},p_T) + D_{\psi}((\lambda,\theta^k),\phi^k); \label{eqn:DivergenceAlgoSimp1} \\
\theta^{k+1} & = & \arginf_{\theta\in\Theta, s.t. (\lambda^{k+1},\theta)\in\Phi} \hat{D}(p_{\lambda^{k+1},\theta},p_T) + D_{\psi}((\lambda^{k+1},\theta),\phi^k).
\label{eqn:DivergenceAlgoSimp2}
\end{eqnarray}
This algorithm corresponds to a directional optimization for recurrence (\ref{eqn:DivergenceAlgo}). We can therefore prove analogously that the estimated divergence between the model and the true density decreases as we proceed with the recurrence.\\

We end the first part of this paper by three remarks:
\begin{itemize}
\item[$\bullet$] Function $\psi$ defining the distance-like proximal term $D_{\psi}$ needs not to be convex as in \cite{Tseng}. As we will see in the convergence proofs, the only properties needed are: $\psi$ is a non negative function defined on $\mathbb{R}_+$ verifying $\psi(t)=0$ iff $t=1$, and $\psi'(t)=0$ iff $t=1$.
\item[$\bullet$] The simplified version is not restricted to mixture models. Indeed, any parametric model, whose vector of parameters can be separated into two independent parts, can be estimated using the simplified version.
\item[$\bullet$] As we will see in the proofs, results on the simplified version can be extended to a further simplified one. In other words, one may even consider an algorithm which attack a lower level of optimization. We may optimize on each class of the mixture model instead of the whole set of parameters. Since the analytic separation is not evident, one should expect some loss of quality as a cost of a less optimization time.
\end{itemize}
The remaining of the paper is devoted entirely to the study of the convergence of the sequences generated by either of the two sets of algorithms (\ref{eqn:DivergenceAlgo}) and (\ref{eqn:DivergenceAlgoSimp1}, \ref{eqn:DivergenceAlgoSimp2}) presented above. A key feature which will be needed in the proofs is the regularity of the objective function $\hat{D}_{\varphi}(p_{\phi},p_T)$. This is the main goal of the following section.
\section{Analytical discussion about the regularity of the estimated divergence}\label{sec:AnalyticalDiscuss}
The estimated divergence in (\ref{eqn:MDPDdef}) or (\ref{eqn:NewDualForm}) has an integral form. Thus, continuity and differentiability can be checked using Lebesgue theorems, see Section \ref{sec:Examples}. However, the dual formula defining the estimator of the $\varphi-$divergence between the true density and the model (\ref{eqn:DivergenceDef}) seems quite complicated. This is basically because of a functional integral and a supremum over it. Continuity and differentiation of the integral is resolved by Lebesgue theorems. We only need that the integrand as well as its partial derivatives to be uniformly bounded with respect to the parameter. However, continuity or differentiability of the supremum is more subtle. Indeed, even if the optimized function is $\mathcal{C}^{\infty}$, it does not imply the continuity of its supremum. Take for example function $f(x,u)=-e^{xu}$. We have:
\[\sup_{x} f(x,u) = \left\{\begin{array}{ccc} -1 & \text{if} & u=0; \\
                                              0 & \text{if} & u\neq 0.
\end{array}\right.\]
On the basis of the theory presented in \cite{Rockafellar} about parametric optimization, we present two ways for studying continuity and differentiability of $\hat{D}_{\varphi}(p_{\phi},p_{\phi_T})$ defined through (\ref{eqn:DivergenceDef}). The first one is the most important because it is easier and demands less mathematical notations. In the first approach, we provide sufficient conditions in order to prove continuity and differentiability almost everywhere of the dual estimator of the divergence. This approach will be used in the study of the convergence of our proximal-point algorithm, see Section \ref{sec:Examples}. The second approach is presented for the sake of completness of the study. We give sufficient conditions which permit to prove the differentiability \emph{everywhere}.\\ 
We recall first the definition of a subgradient of a real valued function $f$.
\begin{definition}[Definition 8.3 in \cite{Rockafellar}]
Consider a function $f:\mathbb{R}^d\rightarrow\bar{\mathbb{R}}$ and a point $\phi^*$ with $f(\phi^*)$ finite. For a vector $v$ in $\mathbb{R}^d$, one says that:
\begin{itemize}
\item[(a)] $v$ is a regular subgradient of $f$ at $\phi^*$, written $v\in\hat{\partial} f(\phi^*)$, if:
\[f(\alpha) \geq f(\phi^*) + <v,\alpha-\phi^*> + o\left(|\alpha-\phi^*|\right);\]
\item[(b)] $v$ is a (general) subgradient of $f$ at $\phi^*$, written $v\in\partial f(\phi^*)$, if there are sequences $\alpha^{n}\rightarrow\phi^*$ with $f(\alpha^n)\rightarrow f(\phi^*)$, and $v^n\in\hat{\partial}f(\alpha^n)$ with $v^n\rightarrow v$.
\end{itemize}
\end{definition}
\subsection{A result of differentiability almost everywhere : Lower-\texorpdfstring{$\mathcal{C}^1$}{TEXT} functions}\label{para:LowerC1}
\begin{definition} [\cite{Rockafellar} Chap 10.] A function $D:\Phi\rightarrow\mathbb{R}$, where $\Phi$ is an open set in $\mathbb{R}^d$, is said to be lower-$\mathcal{C}^1$ on $\Phi$, if on some neighborhood $V$ of each $\phi$ there is a representation
\[D(\phi) = \sup_{\alpha\in T} f(\alpha,\phi)\]
in which the functions $\alpha\mapsto f(\alpha,\phi)$ are of class $\mathcal{C}^1$ on $V$ and the set $T$ is a compact set such that $f(\alpha,\phi)$ and $\nabla_{\phi}f(\alpha,\phi)$ depend continuously not just on $\phi\in \Phi$ but jointly on $(\alpha,\phi)\in T\times V$.
\end{definition}
\noindent In our case, the supremum form is globally defined. Moreover, $T = \Phi$. In case $\Phi$ is bounded, it suffices then to take $T=cl(\Phi)$ the closure of $\Phi$ since $\alpha\mapsto f(\alpha,\phi)$ is continuous. The condition on $T$ to be compact is essential here, and can not be compromised, so that it is necessary to reduce in a way or in another the optimization on $\alpha$ into a compact or at least a bounded set. For example, one may prove that the values of $\alpha\mapsto f(\alpha,\phi)$ near infinity are lower than some value inside $\Phi$ independently of $\phi$.\\
\begin{theorem}[Theorem 10.31 in \cite{Rockafellar}] \label{theo:LowerC1}
Any lower-$\mathcal{C}^1$ function $D$ on an open set $\Phi\subset\mathbb{R}^d$ is both (strictly\footnote{A strictly continuous function $f$ is a local Lipschitz continuous function, i.e. for each $x_0\in\text{int}{\Phi}$, the following limit exists and is finite \[\limsup_{x,x'\rightarrow x_0} \frac{|f(x')-f(x)|}{x'-x}\]}) continuous and continuously differentiable where it is differentiable. Moreover, if $\Delta$ consists of the points where $D$ is differentiable, then $\Phi\setminus\Delta$ is negligible\footnote{A set is called negligible if for every $\varepsilon>0$, there is a family of boxes $\{B_k\}_k$ with $d-$dimensional volumes $\varepsilon_k$ such that $A\subset\cup_{k}{B_k}$ and $\sum_{k}{\varepsilon_k}<\varepsilon$.}.
\end{theorem}
The stated result can be ensured by simple hypotheses on the model $p_{\phi}$ and the function $\varphi$. Unfortunately, since the estimated divergence $\hat{D}_{\varphi}(p_{\phi},p_{\phi_T})$ will not be everywhere differentiable, we can no longer talk about the stationarity of $\hat{D}_{\varphi}(p_{\phi},p_{\phi_T})$ at a limit point of the sequence $\phi^k$ generated for example by (\ref{eqn:DivergenceAlgo}). We therefore, use the notion of subgradients. Indeed, when a function $g$ is not differentiable, a necessary condition for $x_0$ to be a local minimum of $g$ is that $0\in\partial g(x_0)$ and it becomes sufficient whenever $g$ is proper convex\footnote{See \cite{Rockafellar} theorem 10.1.}. Moreover, as $g$ becomes differentiable at $x_0$, then $\nabla g(x_0)\in\partial g(x_0)$ with equality if and only if $g$ is $\mathcal{C}^1$. In other words, proving that $0\in\partial \hat{D}_{\varphi}(p_{\hat{\phi}},p_{\phi_T})$ means that $\hat{\phi}$ is a sort of a \emph{generalized stationary point} of $\phi\mapsto\hat{D}_{\varphi}(p_{\phi},p_{\phi_T})$.\\
We will be studying later on in paragraphs (\ref{Example:CauchyZeroLoc}) and (\ref{Example:DivergenceMixture}) examples where we verify with more details the previous conditions and see the resulting consequences on the sequence $(\phi^k)_k$.
\subsection{A result of everywhere differentiability: Level-bounded functions}\label{para:LevelBoundFun}
\begin{definition}[\cite{Rockafellar} Chap 1.]
A function $f:\mathbb{R}^d\times\mathbb{R}^d\rightarrow\bar{\mathbb{R}}$ with values $f(\alpha,\phi)$ is (upper) level-bounded in $\alpha$ locally uniformly in $\phi$ if for each $\phi_0$ and $a\in\mathbb{R}$ there is a neighborhood $V$ for $\phi_0$ such that the set $\{(\alpha,\phi)|\phi\in V, f(\alpha,\phi)\geq a\}$ is bounded in $\mathbb{R}^d\times\mathbb{R}^d$ for every $a\in\mathbb{R}$.
\end{definition}
For a fixed $\phi$, the level-boundedness property corresponds to having $f(\alpha,\phi)\rightarrow -\infty$ as $\|\alpha\|\rightarrow\infty$. In order to state the main result for this case, let $\phi_0$ be a point at which we need to study continuity and differentiability of $\phi\mapsto\sup_{\alpha}{f(\alpha,\phi)}$. A first result gives sufficient conditions under which the supremum function is continuous. We state it as follows:
\begin{theorem}[\cite{Rockafellar} Theorem 1.17] \label{theo:LevelBoundedCont}
Let $f:\mathbb{R}^n\times\mathbb{R}^m\rightarrow\bar{\mathbb{R}}$ be an upper semicontinuous function. Suppose that $f(\alpha,\phi)$ is level-bounded in $\alpha$ locally uniformly in $\phi$. For function $\phi\mapsto \sup_{\alpha}f(\alpha,\phi)$ to be continuous at $\phi_0$, a sufficient condition is the existence of $\alpha_0\in\argmax_{\alpha}f(\alpha,\phi_0)$ such that $\phi\mapsto f(\alpha_0,\phi)$ is continuous at $\phi_0$.
\end{theorem}
Since in general, we do not know exactly where the supremum will be, one proves the continuity of $\phi\mapsto f(\alpha,\phi)$ for every $\alpha$.\\
A Further result about continuity and differentiability of the supremum function can also be stated. Define, at first, the sets $Y(\phi_0)$ and $Y_{\infty}(\phi_0)$ as follows:
\begin{eqnarray*}
Y(\phi_0) & = & \bigcup_{\alpha\in\argsup_{\beta} f(\beta,\phi_0)} M(\alpha,\phi_0),\quad \text{for } M(\alpha,\phi_0) = \{a|(0,a)\in\partial f(\alpha,\phi_0)\} \\
Y_{\infty}(\phi_0) & = & \bigcup_{\alpha\in\argsup_{\beta} f(\beta,\phi_0)} M_{\infty}(\alpha,\phi_0),\quad \text{for } M_{\infty}(\alpha,\phi_0) = \{a|(0,a)\in\partial^{\infty} f(\alpha,\phi_0)\}
\end{eqnarray*}
where $\partial^{\infty} f$ is the horizon subgradient, see Definition 8.3 (c) in \cite{Rockafellar}. We avoided to mention the definition here in order to keep the text clearer. Furthermore, in the whole chapter, the horizon subgradient will always be equal to the set $\{0\}$.\\
\begin{theorem}[Corollary 10.14 in \cite{Rockafellar}] 
\label{theo:levelbounded}
For a proper upper semicontinuous function $f:\mathbb{R}^d\times\mathbb{R}^d\rightarrow \bar{\mathbb{R}}$ such that $f(\alpha,\phi)$ is level-bounded in $\alpha$ locally uniformly in $\phi$, and for $\phi_0\in$ dom $\sup_{\alpha}f(\alpha,\phi)$:
\begin{itemize}
\item[(a)] If $Y_{\infty}(\phi_0) = \{0\}$, then $\phi\mapsto\sup_{\alpha}f(\alpha,\phi)$ is strictly continuous at $\phi_0$;
\item[(b)] if $Y(\phi_0) = \{a\}$ too, then\footnote{In the statement of the corollary in \cite{Rockafellar}, the supremum function becomes strictly differentiable, but to avoid extra vocabularies, we replaced it with an equivalent property.} $\phi\mapsto\sup_{\alpha}f(\alpha,\phi)$ is $\mathcal{C}^1$ at $\phi_0$ with $\nabla \sup_{\alpha}f(\alpha,\phi) = a$.
\end{itemize}
\end{theorem}
In our examples, $f$ will be a continuous function and even $\mathcal{C}^1(\Phi\times\Phi)$. This implies that $\partial^{\infty} f(\alpha, \phi) = \{0\}$ and $\partial f(\alpha,\phi) = \{\nabla f(\alpha,\phi)\}$, see Exercise 8.8 in \cite{Rockafellar}. Hence, $Y_{\infty}(\phi_0) = \{0\}$ whatever $\phi_0$ in $\Phi$. Moreover $M(\alpha,\phi_0) = \{\nabla_{\phi} f(\alpha,\phi_0)\}$ so that $Y(\phi_0)=\bigcup\{\nabla_{\phi} f(\alpha,\phi_0)\}$ and the union is on the set of suprema of $\alpha\mapsto f(\alpha,\phi_0)$. If $f(\alpha,\phi)$ is level-bounded in $\alpha$ locally uniformly in $\phi$, then the supremum function becomes strictly continuous. Moreover, if the function $f$ has the same gradient with respect to $\phi$ for all the suprema of $\alpha\mapsto f(\alpha,\phi)$, then $\sup_{\alpha} f(\alpha,\phi)$ becomes continuously differentiable. This is for example the case when function $\alpha\mapsto f(\alpha,\phi)$ has a unique global supremum for a fixed $\phi$, which is for example the case of a strictly concave function (with respect to $\alpha$ for a fixed $\phi$).\\

\section{Some convergence properties of \texorpdfstring{$\phi^k$}{phik}}\label{sec:Proofs}
We adapt the ideas given in \cite{Tseng} to develop a suitable proof for our proximal algorithm. We present some propositions which show how according to some possible situations one may prove convergence of the algorithms defined by recurrences (\ref{eqn:DivergenceAlgo}) and (\ref{eqn:DivergenceAlgoSimp1}, \ref{eqn:DivergenceAlgoSimp2}). Let $\phi^0=(\lambda^0,\theta^0)$ be a given initialization for the parameters, and define the following set
\begin{equation}
\Phi^0 = \{\phi\in\Phi: \hat{D}(p_{\phi},p_{\phi_T})\leq \hat{D}(\phi^0,\phi_T)\}.
\label{eqn:SetPhis0}
\end{equation} 
We suppose that $\Phi^0$ is a subset of $int(\Phi)$. The idea of defining such set in this context is inherited from the paper of \cite{Wu} which provided the first \emph{correct proof} of convergence for the EM algorithm. Before going any further, we recall the following definition of a (generalized) stationary point.\\ 
\begin{definition}
Let $f:\mathbb{R}^d\rightarrow\mathbb{R}$ be a real valued function. If $f$ is differentiable at a point $\phi^*$ such that $\nabla f(\phi^*)=0$, we then say that $\phi^*$ is a stationary point of $f$. If $f$ is not differentiable at $\phi^*$ but the subgradient of $f$at $\phi^*$, say $\partial f(\phi^*)$, exists such that $0\in\partial f(\phi^*)$, then $\phi^*$ is called a generalized stationary point of $f$.
\end{definition}
We will be using the following assumptions which will be checked in several examples later on.
\begin{itemize}
\item[A0.] Functions $\phi\mapsto\hat{D}(p_{\phi}|p_{\phi_T}), D_{\psi}$ are lower semicontinuous;
\item[A1.] Functions $\phi\mapsto\hat{D}(p_{\phi}|p_{\phi_T}), D_{\psi}$ and $\nabla_1 D_{\psi}$ are defined and continuous on, respectively, $\Phi, \Phi\times\Phi$ and $\Phi\times\Phi$;
\item[AC.] $\nabla \hat{D}(p_{\phi}|p_{\phi_T})$ is defined and continuous on $\Phi$;
\item[A2.] $\Phi^0$ is a compact subset of int($\Phi$);
\item[A3.] $D_{\psi}(\phi,\bar{\phi})>0$ for all $\bar{\phi}\neq \phi \in \Phi$.
\end{itemize}
Recall also the assumptions on functions $h_i$ defining $D_{\psi}$. We suppose that $h_i(x|\phi)>0, dx-a.e.$, and $\psi(t)=0$ iff $t=1$. Besides $\psi'(t)=0$ iff $t = 1$.\\
Concerning assumptions A1 and AC, we have previously discussed the analytical properties of $\hat{D}(p_{\phi}|p_{\phi_T})$ in Section \ref{sec:AnalyticalDiscuss}. In what concerns $D_{\psi}$, continuity and differentiability can be obtained merely by fulfilling Lebesgue theorems conditions. For example, if $h_i(x,\phi)$ is continuous and bounded uniformly away from 0 independently of $\phi$, then continuity is guaranteed as soon as $\psi$ is continuous. If we also suppose that $\nabla_{\phi} h_i(x,\phi)$ exists, is continuous and is uniformly bounded independently of $\phi$, then as soon as $\psi$ is continuously differentiable, $D_{\psi}$ becomes continuously differentiable. For assumption A2, there is no universal method. Still, in all the examples that will be discussed later, we use the fact that the inverse image of a closed set by a continuous function is closed. Boundedness is usually ensured using a \emph{suitable} choice of $\phi^0$. Finally, assumption A3 is checked using Lemma 2 proved in \cite{Tseng} which we restate here.
\begin{lemma}[Lemma 2 in \cite{Tseng}] \label{Lem:Tseng}
Suppose $\psi$ to be a continuous nonnegative function such that $\psi(t)=0$ iff $t=1$. For any $\phi$ and $\phi'$ in $\Phi$, if $h_i(x|\phi)\neq h_i(x|\phi')$ for some $i\in\{1,\cdots,n\}$ and some $x\in int(X)$ at which both $h_i(.|\phi)$ and $h_i(.|\phi')$ are continuous, then $D_{\psi}(\phi,\phi')>0$.
\end{lemma}
\noindent In section (\ref{sec:Examples}), we present three different examples; a two-component Gaussian mixture, a two-component Weibull mixture and a Cauchy model. We will see that the Cauchy example verifies assumption A3. However, the Gaussian mixture does not seem to verify it. Indeed, the same fact stays true for any mixture of the exponential family.\\
We start by providing some general facts about the sequence $(\phi^k)_k$ and its existence. We also prove convergence of the sequence $(\hat{D}(p_{\phi^k}|p_{\phi_T}))_k$.
\begin{remark}
All results concerning algorithm (\ref{eqn:DivergenceAlgo}) are proved even when assumption AC is not fulfilled. We give proofs using the subgradient of the estimated $\varphi-$divergence. In the case of the two-step algorithm (\ref{eqn:DivergenceAlgoSimp1}, \ref{eqn:DivergenceAlgoSimp2}), it was not possible and thus remains an open problem. The difficulty resides in manipulating the \emph{partial} subgradients with respect to $\lambda$ and $\theta$ which cannot be handled in a similar way to the partial derivatives.
\end{remark}
\begin{remark}
Convergence properties are proved without using the special form of the estimated $\varphi-$divergence. Thus, our theoretical approach applies to any optimization problem whose objective is to minimize a function $\phi\mapsto D(\phi)$. For example, our approach can be applied on density power divergences (\ref{eqn:MDPDdef}), (kernel-based) MD$\varphi$DE (\ref{eqn:MDphiDEClassique},\ref{eqn:NewMDphiDE}), Bregman divergences, S-divergences (\cite{GoshSDivergence}), R\'enyi pseudodistances (see for example \cite{TomaAubin}), etc.
\end{remark}
The proofs of Propositions \ref{prop:DecreaseDphi}, \ref{prop:StationaryPhiDiff} and \ref{prop:PhiDiffConverge} are only given for the two-step algorithm (\ref{eqn:DivergenceAlgoSimp1}, \ref{eqn:DivergenceAlgoSimp2}). The proofs of the case of algorithm (\ref{eqn:DivergenceAlgo}) are direct adaptations of Theorem 1 and Lemme 1 in \cite{Tseng} for the case of the likelihood function, see also \cite{DiaaBroniaProximalEntropy}. The proofs when assumption AC is not fulfilled can be found in \cite{DiaaBroniaProximalEntropy} with $\hat{D}_{\varphi}$ instead of $\hat{D}$.
\begin{proposition}
\label{prop:DecreaseDphi}
We assume that recurrences (\ref{eqn:DivergenceAlgo}) and (\ref{eqn:DivergenceAlgoSimp1}, \ref{eqn:DivergenceAlgoSimp2}) are well defined in $\Phi$. For both algorithms, the sequence $(\phi^{k})_k$ verifies the following properties:
\begin{itemize}
\item[(a)] $D_{\varphi}(p_{\phi^{k+1}}|p_T)\leq D_{\varphi}(p_{\phi^k}|p_T)$;
\item[(b)] $\forall k, \phi^k \in \Phi^0$;
\item[(c)] Suppose that assumptions A0 and A2 are fulfilled, then the sequence $(\phi^k)_k$ is defined and bounded. Moreover, the sequence $\left(\hat{D}(\phi^k|\phi_T)\right)_k$ converges.
\end{itemize}
\end{proposition}
\begin{IEEEproof}
\underline{We prove $(a)$}. \textbf{For the two-step algorithm} defined by (\ref{eqn:DivergenceAlgoSimp1}, \ref{eqn:DivergenceAlgoSimp2}), recurrence (\ref{eqn:DivergenceAlgoSimp1}) and the definition of the arginf give:
\begin{eqnarray}
\hat{D}(p_{\lambda^{k+1},\theta^k},p_T) + D_{\psi}((\lambda^{k+1},\theta^k),\phi^k) & \leq & \hat{D}(p_{\lambda^k,\theta^k},p_T) + D_{\psi}((\lambda^k,\theta^k),\phi^k) \nonumber\\
					& \leq & \hat{D}(p_{\lambda^k,\theta^k},p_T).
\label{eqn:PartialDecrease1}					
\end{eqnarray}
The second inequality is obtained using the fact that $D_{\psi}(\phi,\phi)=0$. Using recurrence (\ref{eqn:DivergenceAlgoSimp2}), we get:
\begin{eqnarray}
\hat{D}(p_{\lambda^{k+1},\theta^k},p_T) + D_{\psi}((\lambda^{k+1},\theta^k),\phi^k) & \geq & \hat{D}(p_{\lambda^{k+1},\theta^{k+1}},p_T) + D_{\psi}((\lambda^{k+1},\theta^{k+1}),\phi^k) \\
  					& \geq & \hat{D}(p_{\lambda^{k+1},\theta^{k+1}},p_T).
\label{eqn:PartialDecrease2}
\end{eqnarray}
The second inequality is obtained using the fact that $D(\phi|\phi')\geq 0$. The conclusion is reached by combining the two inequalities (\ref{eqn:PartialDecrease1}) and (\ref{eqn:PartialDecrease2}).\\
\underline{We prove $(b)$}. Using the decreasing property previously proved in (a), we have by recurrence $\forall k, \hat{D}(p_{\phi^{k+1}},p_T)\leq \hat{D}(p_{\phi^k},p_T)\leq\cdots \leq \hat{D}(p_{\phi^0},p_T)$. The result follows for both algorithms directly by definition of $\Phi^0$.\\
\underline{We prove $(c)$}. By induction on $k$. For $k=0$, clearly $\phi^0 = (\lambda^0,\theta^0)$ is well defined (a choice we make\footnote{The choice of the initial point of the sequence may influence the convergence of the sequence. See the example of the Gaussian mixture in paragraph (\ref{Example:GaussMix}).}). Suppose for some $k\geq 0$ that $\phi^k = (\lambda^k,\theta^k)$ exists. \textbf{For the two-step algorithm} defined by (\ref{eqn:DivergenceAlgoSimp1},\ref{eqn:DivergenceAlgoSimp2}). The infimum in (\ref{eqn:DivergenceAlgoSimp1}) can be calculated on $\lambda$'s such that $(\lambda,\theta^k)\in\Phi^0$. Indeed, suppose there exists a $\lambda$ at which the value of the optimized function is less than its value at $\lambda^k$, i.e. $\hat{D}(p_{\lambda,\theta^k},p_T) + D_{\psi}((\lambda,\theta^k),\phi^k) \leq \hat{D}(p_{\lambda^k,\theta^k},p_T) + D_{\psi}((\lambda^k,\theta^k),\phi^k)$. We have:
\begin{eqnarray*}
\hat{D}(p_{\lambda,\theta^k},p_T) & \leq & \hat{D}(p_{\lambda,\theta^k},p_T) + D_{\psi}((\lambda,\theta^k),\phi^k) \\
& \leq & \hat{D}(p_{\lambda^k,\theta^k},p_T) + D_{\psi}((\lambda^k,\theta^k),\phi^k) \\
  & \leq & \hat{D}(p_{\lambda^k,\theta^k},p_T) \\
  & \leq & \hat{D}(p_{\phi^0},p_T).
\end{eqnarray*}
This means that $(\lambda,\theta^k)\in\Phi^0$ and that the infimum needs not to be calculated for all values of $\lambda$ in $\Phi$, and can be restrained onto values which verify $(\lambda,\theta^k)\in\Phi^0$.\\
Define now $\Lambda_k = \{\lambda\in [0,1]^s| (\lambda,\theta^k)\in\Phi^0\}$. First of all, $\lambda^k\in\Lambda_k$ since $(\lambda^k,\theta^k)\in\Phi^0$. Therefore, $\Lambda_k$ is not empty. Moreover, it is compact. Indeed, let $(\lambda^l)_l$ be a sequence of elements of $\Lambda_k$, then the sequence $((\lambda^l,\theta^k))_l$ is a sequence of elements of $\Phi^0$. By compactness of $\Phi^0$, there exists a subsequence which converges in $\Phi^0$ to an element of the form $(\lambda^{\infty},\theta^k)$ which clearly belongs to $\Lambda_k$. This proves that $\Lambda_k$ is compact. Finally, since by assumption A0, the optimized function is lower semicontinuous so that it attains its infimum on the compact set $\Lambda_k$. We may now define $\lambda^{k+1}$ as any vector verifying this infimum.\\
The second part of the proof treats the definition of $\theta^{k+1}$. Let $\theta$ be any vector such that $(\lambda^{k+1},\theta)\in\Phi$ and at which the value of the optimized function in (\ref{eqn:DivergenceAlgoSimp2}) is less than its value at $\phi^k$. We have
\begin{eqnarray*}
\hat{D}(p_{\lambda^{k+1},\theta},p_T) & \leq & \hat{D}(p_{\lambda^{k+1},\theta},p_T) + D_{\psi}((\lambda^{k+1},\theta),\phi^k) \\
& \leq & \hat{D}(p_{\lambda^{k+1},\theta^k},p_T) + D_{\psi}((\lambda^{k+1},\theta^k),\phi^k) \\
& \leq & \hat{D}(p_{\lambda^k,\theta^k},p_T) + D_{\psi}((\lambda^k,\theta^k),\phi^k) \\
& \leq & \hat{D}(p_{\lambda^k,\theta^k},p_T) \\
& \leq & \hat{D}(p_{\phi^0},p_T)
\end{eqnarray*}
The third line comes from the previous definition of $\lambda^{k+1}$ as an infimum of (\ref{eqn:DivergenceAlgoSimp1}). This means that $(\lambda^{k+1},\theta)\in\Phi^0$, and that the infimum in (\ref{eqn:DivergenceAlgoSimp2}) can be calculated with respect to values $\theta$ which verifies $(\theta,\lambda^{k+1})\in\Phi^0$. Define now $\Theta_k = \{\theta\in\mathbb{R}^{d-s}| (\lambda^{k+1},\theta)\in\Phi^0\}$. One can prove analogously to $\Lambda_k$, that it is compact. The optimized function in (\ref{eqn:DivergenceAlgoSimp2}) is, by assumption A0, lower semicontinuous so that its infimum is attained on the compact $\Theta_k$. We may now define $\theta^{k+1}$ as any vector verifying this infimum.\\
Convergence of the sequence $(\hat{D}(p_{\phi^k},p_T))_k$ in both algorithms comes from the fact that it is nonincreasing and bounded. It is nonincreasing by virtue of (a). Boundedness comes from the lower semicontinuity of $\phi\mapsto\hat{D}(p_{\phi},p_T)$. Indeed, $\forall k, \hat{D}(p_{\phi^k},p_T) \geq \inf_{\phi\in\Phi^0}\hat{D}(p_{\phi},p_T)$. The infimum of a proper lower semicontinuous function on a compact set exists and is attained on this set. Hence, the quantity $\inf_{\phi\in\Phi^0}\hat{D}(p_{\phi},p_T)$ exists and is finite. This ends the proof.
\end{IEEEproof}
\noindent The interest of Proposition \ref{prop:DecreaseDphi} is that the objective function is ensured, under mild assumptions, to decrease alongside the sequence $(\phi^k)_k$. This permits to build a stop criterion for the algorithm since in general there is no guarantee that the whole sequence $(\phi^k)_k$ converges. It may also continue to fluctuate in a neighborhood of an optimum. The following result provides a first characterization about the properties of the limit of the sequence $(\phi^k)_k$ as (generalized) a stationary point of the estimated $\varphi-$divergence.
\begin{proposition}
\label{prop:StationaryPhiDiff}
Suppose that A1 is verified, and assume that $\Phi^0$ is closed and $\{\phi^{k+1}-\phi^k\}\rightarrow 0$.
\begin{itemize}
\item[(a)] For both algorithms (\ref{eqn:DivergenceAlgo}) and (\ref{eqn:DivergenceAlgoSimp1},\ref{eqn:DivergenceAlgoSimp2}), if AC is verified, then the limit of every convergent subsequence is a stationary point of $\hat{D}(.|p_T)$;
\item[(b)] For the first algorithm (\ref{eqn:DivergenceAlgo}), if $\hat{D}(.|p_T)$ is not differentiable,  then the limit of every convergent subsequence is a "generalized" stationary point of $\hat{D}(.|p_T)$, i.e. zero belongs to the subgradient of $\hat{D}(.|p_T)$ calculated at the limit point;
\end{itemize}
\end{proposition}
\begin{IEEEproof} 
\underline{We prove $(a)$}. Let $(\phi^{n_k})_k$ be a convergent subsequence of $(\phi^k)_k$ which converges to $\phi^{\infty}$. First, $\phi^{\infty} \in \Phi^0$, because $\Phi^0$ is closed and the subsequence $(\phi^{n_k})$ is a sequence of elements of $\Phi^0$ (proved in Proposition \ref{prop:DecreaseDphi}.b).\\
Let's show now that the subsequence $(\phi^{n_k+1})$ also converges to $\phi^{\infty}$. We simply have:
\begin{eqnarray*}
\|\phi^{n_k+1} - \phi^{\infty}\| & \leq & \|\phi^{n_k} - \phi^{\infty}\| + \|\phi^{n_k+1} - \phi^{n_k}\|
\end{eqnarray*}
Since $\phi^{k+1} - \phi^k \rightarrow 0$ and $\phi^{n_k} \rightarrow \phi^{\infty}$, we conclude that $\phi^{n_k+1} \rightarrow \phi^{\infty}$.\\
\textbf{For the two-step algorithm (\ref{eqn:DivergenceAlgoSimp1},\ref{eqn:DivergenceAlgoSimp2})}, by definition of $\lambda^{n_k+1}$ and $\theta^{n_k+1}$, they verify the infimum respectively in recurrences (\ref{eqn:DivergenceAlgoSimp1}) and (\ref{eqn:DivergenceAlgoSimp2}). Therefore, the gradient of the optimized function is zero for each step. In other words:
\begin{eqnarray*}
\nabla_{\lambda} \hat{D}(p_{\lambda^{n_k+1},\theta^{n_k}},p_T) + \nabla_{\lambda} D_{\psi}((\lambda^{n_k+1},\theta^{n_k}),\phi^{n_k}) & = & 0 \\
\nabla_{\theta} \hat{D}(p_{\lambda^{n_k+1},\theta^{n_k+1}},p_T) + \nabla_{\theta} D_{\psi}((\lambda^{n_k+1},\theta^{n_k+1}),\phi^{n_k}) & = & 0
\end{eqnarray*}
Since both $(\phi^{n_k+1})$ and $(\phi^{n_k})$ converge to the same limit $\phi^{\infty}$, then setting $\phi^{\infty} = (\lambda^{\infty},\theta^{\infty})$, we get $\lambda^{n_k+1}$ and $\lambda^{n_k}$ tends to $\lambda^{\infty}$. We also have $\theta^{n_k+1}$ and $\theta^{n_k}$ tends to $\theta^{\infty}$. The continuity of the two gradients (assumptions A1 and AC) implies that:
\begin{eqnarray*}
\nabla_{\lambda} \hat{D}(p_{\lambda^{\infty},\theta^{\infty}},p_T) + \nabla_{\lambda} D_{\psi}((\lambda^{\infty},\theta^{\infty}),\phi^{\infty}) & = & 0 \\
\nabla_{\theta} \hat{D}(p_{\lambda^{\infty},\theta^{\infty}},p_T) + \nabla_{\theta} D_{\psi}((\lambda^{\infty},\theta^{\infty}),\phi^{\infty}) & = & 0
\end{eqnarray*}
However, $\nabla D_{\psi}(\phi,\phi) = 0$, so that $\nabla_{\lambda} \hat{D}(p_{\phi^{\infty}},p_T)=0$ and $\nabla_{\theta} \hat{D}(p_{\phi^{\infty}},p_T)=0$. Hence $\nabla \hat{D}(p_{\phi^{\infty}},p_T)=0$.\\
\underline{We prove (b)}. See the proof of Proposition 2-b in \cite{DiaaBroniaProximalEntropy}.
\end{IEEEproof}
\begin{proposition}
\label{prop:PhiDiffConverge}
For both algorithms defined by (\ref{eqn:DivergenceAlgo}) and (\ref{eqn:DivergenceAlgoSimp1},\ref{eqn:DivergenceAlgoSimp2}), assume A1, A2 and A3 verified, then $\{\phi^{k+1}-\phi^k\}\rightarrow 0$. Thus, by Proposition \ref{prop:StationaryPhiDiff} (according to whether AC is verified or not) implies that any limit point of the sequence $\phi^k$ is a (generalized)\footnote{The case where AC is not verified is only proved for the first algorithm (\ref{eqn:DivergenceAlgo})} stationary point of $\hat{D}(.|p_T)$.
\end{proposition}
\begin{IEEEproof}
The arguments presented are the same for both algorithms (\ref{eqn:DivergenceAlgo}) and (\ref{eqn:DivergenceAlgoSimp1},\ref{eqn:DivergenceAlgoSimp2}). By contradiction, let's suppose that $\phi^{k+1}-\phi^k$ does not converge to 0. There exists a subsequence such that $\|\phi^{N_0(k)+1}-\phi^{N_0(k)}\| > \varepsilon,\; \forall k\geq k_0$. Since $(\phi^k)_k$ belongs to the compact set $\Phi^0$, there exists a convergent subsequence $(\phi^{N_1\circ N_0(k)})_k$ such that $\phi^{N_1\circ N_0(k)}\rightarrow \bar{\phi}$. The sequence $(\phi^{N_1\circ N_0(k)+1})_k$ belongs to the compact set $\Phi^0$, therefore we can extract a further subsequence $(\phi^{N_2\circ N_1\circ N_0(k)+1})_k$ such that $\phi^{N_2\circ N_1\circ N_0(k)+1}\rightarrow \tilde{\phi}$. Besides $\hat{\phi}\neq \tilde{\phi}$. Finally since the sequence $(\phi^{N_1\circ N_0(k)})_k$ is convergent, a further subsequence also converges to the same limit $\bar{\phi}$. We have proved the existence of a subsequence of $(\phi^k)_k$ such that $\phi^{N(k)+1}-\phi^{N(k)}$ does not converge to 0 and such that $\phi^{N(k)+1} \rightarrow \tilde{\phi}$, $\phi^{N(k)} \rightarrow \bar{\phi}$ with $\bar{\phi} \neq \tilde{\phi}$.\\
The real sequence $\hat{D}(p_{\phi^k},p_T)_k$ converges as proved in Proposition \ref{prop:DecreaseDphi}-c. As a result, both sequences $\hat{D}(p_{\phi^{N(k)+1}},p_T)$ and $\hat{D}(p_{\phi^{N(k)}},p_T)$ converge to the same limit being subsequences of the same convergent sequence. In the proof of Proposition \ref{prop:DecreaseDphi}, we can deduce the following inequality:
\begin{equation}
\hat{D}(p_{\lambda^{k+1},\theta^{k+1}},p_T) + D_{\psi}((\lambda^{k+1},\theta^{k+1}),\phi^k) \leq \hat{D}(p_{\lambda^k,\theta^k},p_T)
\label{eqn:DivergenceDecreaseSeq}
\end{equation}
which is also verified to any substitution of $k$ by $N(k)$. By passing to the limit on k, we get $D_{\psi}(\tilde{\phi},\bar{\phi}) \leq 0$. However, the distance-like function $D_{\psi}$ is positive, so that it becomes zero. Using assumption A3, $D_{\psi}(\tilde{\phi},\bar{\phi}) = 0$ implies that $\tilde{\phi} = \bar{\phi}$. This contradicts the hypothesis that $\phi^{k+1}-\phi^k$ does not converge to 0.\\
The second part of the proposition is a direct result of Proposition \ref{prop:StationaryPhiDiff}.
\end{IEEEproof}
We can go further in exploring the properties of the sequence $(\phi^k)_k$, but we need to impose more assumptions. The following corollary provides a convergence result of the \emph{whole} sequence and not only some subsequence. The convergence is also towards a local minimum as soon as the estimated divergence is locally strictly convex.
\begin{corollary} 
Under assumptions of Proposition \ref{prop:PhiDiffConverge}, the set of accumulation points of $(\phi^k)_k$ is a connected compact set. Moreover, if $\hat{D}(p_{\phi},p_T)$ is strictly convex in a neighborhood of a limit point\footnote{This assumption can be replaced by local strict convexity since \emph{a priori}, we have no idea where might find a limit point of the sequence $(\phi^k)_k$.} of the sequence $(\phi^k)_k$, then the whole sequence $(\phi^k)_k$ converges to a local minimum of $\hat{D}(p_{\phi},p_T)$.
\end{corollary}
\begin{IEEEproof}
The proof is based on Theorem 28.1 in \cite{Ostrowski}, see \cite{DiaaBroniaProximalEntropy}.
\end{IEEEproof}
\noindent Proposition \ref{prop:PhiDiffConverge} although provides a general solution to prove that $\{\phi^{k+1}-\phi^k\}\rightarrow 0$, the identifiability assumption over the proximal term is hard to be fulfilled. It is not verified in the most simple mixtures such as a two component Gaussian mixture, see Section (\ref{Example:GaussMix}).\\
This was the reason behind our next result. We prove that we do not need to assume identifiability of the proximal term in order to prove that any convergent subsequence of $(\phi^k)_k$ is a (generalized) stationary point of the estimated $\varphi-$divergence.\\
A similar idea was employed in \cite{ChretienHeroProxGener} who studied a proximal algorithm for the log-likelihood function with a relaxation parameter\footnote{A sequence of decreasing positive numbers multiplied by the proximal term.}. Their work however requires that the log-likelihood has $-\infty$ limit as $\|\phi\|\rightarrow\infty$ which is simply not verified on several mixture models (e.g. the Gaussian mixture model). Our result treat the problem from another approach based on the introduction of the set $\Phi^0$. The following result was already presented in the case of $\varphi-$divergences by \cite{DiaaBroniaProximalEntropy}, but since this result is still new, we prefer to rewrite the proof in the context of our paper.\\
\begin{proposition}
\label{prop:NewRes}
Assume A1, AC and A2 verified. For the algorithm defined by (\ref{eqn:DivergenceAlgo}), any convergent subsequence converges to a stationary point of the objective function $\phi\rightarrow\hat{D}(p_{\phi},p_T)$. If AC is dropped, then 0  belongs to the subgradient of $\phi\mapsto\hat{D}(p_{\phi},p_T)$ at the limit point.
\end{proposition}
\begin{IEEEproof} 
If $(\phi^k)_k$ converges to, say, $\phi^{\infty}$, the result falls simply from Proposition \ref{prop:StationaryPhiDiff}.\\
If $(\phi^k)_k$ does not converge. Since $\Phi^0$ is compact and $\forall k, \phi^k\in\Phi^0$ (proved in Proposition \ref{prop:DecreaseDphi}), there exists a subsequence $(\phi^{N_0(k)})_k$ such that $\phi^{N_0(k)}\rightarrow\tilde{\phi}$. Let's take the subsequence $(\phi^{N_0(k)-1})_k$. This subsequence does not necessarily converge; still it is contained in the compact $\Phi^0$, so that we can extract a further subsequence $(\phi^{N_1\circ N_0(k)-1})_k$ which converges to, say, $\bar{\phi}$. Now, the subsequence $(\phi^{N_1\circ N_0(k)})_k$ converges to $\tilde{\phi}$, because it is a subsequence of $(\phi^{N_0(k)})_k$. We have proved until now the existence of two convergent subsequences $\phi^{N(k)-1}$ and $\phi^{N(k)}$ with \emph{a priori} different limits. For simplicity and without any loss of generality, we will consider these subsequences to be $\phi^k$ and $\phi^{k+1}$ respectively.\\
Conserving previous notations, suppose that $\phi^{k+1}\rightarrow \tilde{\phi}$ and $\phi^{k}\rightarrow \bar{\phi}$. We use again inequality (\ref{eqn:DivergenceDecreaseSeq}):
\[\hat{D}(p_{\phi^{k+1}},p_T) + D_{\psi}(\phi^{k+1},\phi^k) \leq \hat{D}(p_{\lambda^k,\theta^k},p_T)\]
By taking the limits of the two parts of the inequality as $k$ tends to infinity, and using the continuity of the two functions, we have 
\[\hat{D}(p_{\tilde{\phi}},p_T) + D_{\psi}(\tilde{\phi},\bar{\phi}) \leq \hat{D}(p_{\bar{\phi}},p_T)\]
Recall that under A1-2, the sequence $\left(\hat{D}(p_{\phi^k},p_T)\right)_k$ converges, so that it has the same limit for any subsequence, i.e. $\hat{D}(p_{\tilde{\phi}},p_T) = \hat{D}(p_{\bar{\phi}},p_T)$. We also use the fact that the distance-like function $D_{\psi}$ is nonnegative to deduce that $D_{\psi}(\tilde{\phi},\bar{\phi}) = 0$. Looking closely at the definition of this divergence (\ref{eqn:DivergenceClasses}), we get that if the sum is zero, then each term is also zero since all terms are nonnegative. This means that:
\[\forall i\in\{1,\cdots,n\}, \quad \int_{\mathcal{X}}{\psi\left(\frac{h_i(x|\tilde{\phi})}{h_i(x|\bar{\phi})}\right)h_i(x|\bar{\phi})dx} = 0\]
The integrands are nonnegative functions, so they vanish almost ever where with respect to the measure $dx$ defined on the space of labels.
\[\forall i\in\{1,\cdots,n\}, \quad \psi\left(\frac{h_i(x|\tilde{\phi})}{h_i(x|\bar{\phi})}\right)h_i(x|\bar{\phi}) = 0\quad dx-a.e.\]
The conditional densities $h_i$ are supposed to be positive\footnote{In the case of two Gaussian (or more generally exponential) components, this is justified by virtue of a suitable choice of the initial condition.}, i.e. $ h_i(x|\bar{\phi})>0, dx-a.e.$. Hence, $\psi\left(\frac{h_i(x|\tilde{\phi})}{h_i(x|\bar{\phi})}\right) = 0, dx-a.e.$. On the other hand, $\psi$ is chosen in a way that $\psi(z)=0$ iff $z=1$, therefore :
\begin{equation}
\forall i\in\{1,\cdots,n\},\quad h_i(x|\tilde{\phi}) = h_i(x|\bar{\phi}) \quad dx-a.e.
\label{eqn:ProportionsEquality}
\end{equation}
Since $\phi^{k+1}$ is, by definition, an infimum of $\phi\mapsto\hat{D}(p_{\phi},p_T) + D_{\psi}(\phi,\phi^k)$, then the gradient of this function is zero on $\phi^{k+1}$. It results that:
\[\nabla \hat{D}(p_{\phi^{k+1}},p_T) + \nabla D_{\psi}(\phi^{k+1},\phi^k) = 0,\quad \forall k\]
Taking the limit on $k$, and using the continuity of the derivatives, we get that:
\begin{equation}
\nabla \hat{D}(p_{\tilde{\phi}},p_T) + \nabla D_{\psi}(\tilde{\phi},\bar{\phi}) = 0
\label{eqn:GradientLimit}
\end{equation}
Let's write explicitly the gradient of the second divergence:
\[\nabla D_{\psi}(\tilde{\phi},\bar{\phi}) = \sum_{i=1}^n\int_{\mathcal{X}}{\frac{\nabla h_i(x|\tilde{\phi})}{h_i(x|\bar{\phi})}\psi'\left(\frac{h_i(x|\tilde{\phi})}{h_i(x|\bar{\phi})}\right)h_i(x|\bar{\phi})}\]
We use now the identities (\ref{eqn:ProportionsEquality}), and the fact that $\psi'(1)=0$, to deduce that:
\[\nabla D_{\psi}(\tilde{\phi},\bar{\phi}) = 0 \]
This entails using (\ref{eqn:GradientLimit}) that $\nabla \hat{D}(p_{\tilde{\phi}},p_T) = 0$.\\ 
Comparing the proved result with the notation considered at the beginning of the proof, we have proved that the limit of the subsequence $(\phi^{N_1\circ N_0(k)})_k$ is a stationary point of the objective function. Therefore, The final step is to deduce the same result on the original convergent subsequence $(\phi^{N_0(k)})_k$. This is simply due to the fact that $(\phi^{N_1\circ N_0(k)})_k$ is a subsequence of the convergent sequence $(\phi^{N_0(k)})_k$, hence they have the same limit. \\
\textbf{When assumption AC is dropped,} the optimality condition in (\ref{eqn:DivergenceAlgo}) implies :
\[-\nabla D_{\psi}(\phi^{k+1},\phi^k) \in \partial \hat{D}(p_{\phi^{k+1}},p_T)\quad \forall k\]
Function $\phi\mapsto\hat{D}(p_{\phi},p_T)$ is continuous, hence its subgradient is outer semicontinuous and:
\begin{equation}
\limsup_{\phi^{k+1}\rightarrow\phi^{\infty}} \partial \hat{D}(p_{\phi^{k+1}},p_T)\subset \partial \hat{D}(p_{\tilde{\phi}},p_T)
\label{eqn:OSCInclusion}
\end{equation}
By definition of limsup:
\[\limsup_{\phi\rightarrow\phi^{\infty}} \partial \hat{D}(p_{\phi},p_T) = \left\{ u|\exists \phi^k\rightarrow\phi^{\infty},\exists u^k\rightarrow u \text{ with } u^k\in \partial \hat{D}(p_{\phi^k},p_T)\right\}\]
In our scenario, $\phi = \phi^{k+1}$, $\phi^k = \phi^{k+1}$, $u = 0$ and $u^k = \nabla_1 D_{\psi}(\phi^{k+1},\phi^{k})$. We have proved above in this proof that $\nabla_1 D_{\psi}(\tilde{\phi},\bar{\phi}) = 0$ using only convergence of $(\hat{D}(p_{\phi^k},p_T))_k$, inequality (\ref{eqn:DivergenceDecreaseSeq}) and some properties of $D_{\psi}$. Assumption AC was not needed. Hence, $u^k\rightarrow 0$. This proves that, $u = 0\in\limsup_{\phi^{k+1}\rightarrow\phi^{\infty}} \partial \hat{D}(p_{\phi^{n_k+1}},p_T)$. Finally, using the inclusion (\ref{eqn:OSCInclusion}), we get our result:
\[0\in \partial \hat{D}(p_{\tilde{\phi}},p_T)\]
\end{IEEEproof}
\noindent We could not perform the same idea on the two-step algorithm (\ref{eqn:DivergenceAlgoSimp1},\ref{eqn:DivergenceAlgoSimp2}) without assuming that the difference between two consecutive terms of either the sequence of weights $(\lambda^k)_k$ or the sequence of form parameters $(\theta^k)_k$ converges to zero. Besides, when assumption AC is dropped, the proof becomes very complicated because we are obliged to work with partial subgradients. The problem is that the subgradient is a set-valued function and if zero belongs to both the partial subgradients with respect to $\lambda$ and $\theta$ of the objective function, there is no guarantee that it belongs to the "whole" subgradient of the objective function. Hence, we do not have the elements of proof for such result for the time being.
\begin{proposition}
Assume A1 and A2 verified. For the algorithm defined by (\ref{eqn:DivergenceAlgoSimp1},\ref{eqn:DivergenceAlgoSimp2}). If $\|\theta^{k+1} - \theta^k\|\rightarrow 0$, then any convergent subsequence $(\phi^{N(k)})_k$ converges to a stationary point of the objective function $\phi\rightarrow\hat{D}(p_{\phi},p_T)$.
\end{proposition}
\begin{IEEEproof}
\underline{We prove (a)}. We use the same lines from the previous proof to deduce the existence of two convergent subsequences $\phi^{N(k)-1}$ and $\phi^{N(k)}$ with \emph{a priori} different limits. For simplicity and without any loss of generality, we will consider these subsequences to be $\phi^k$ and $\phi^{k+1}$ respectively. Suppose that $\phi^k\rightarrow\bar{\phi} = (\bar{\lambda},\bar{\theta})$ and $\phi^{k+1}\rightarrow\tilde{\phi} = (\tilde{\lambda},\tilde{\theta})$.\\
We first use inequality (\ref{eqn:DivergenceDecreaseSeq}) as in the previous proposition, the convergence of the sequence $(\hat{D}(p_{\lambda^k,\theta^k},p_T))_k$ and some basic properties of $D_{\psi}$ to deduce that:
\begin{equation}
\forall i\in\{1,\cdots,n\},\quad h_i(x|\tilde{\phi}) = h_i(x|\bar{\phi}) \quad dx-a.e.
\label{eqn:ProportionsEqualityAll}
\end{equation}
Let's calculate the gradient of the objective function with respect to $\lambda$ and $\theta$ separately at the limit of $(\phi^{k+1})_k$. By definition of $\theta^{k+1}$ as an arginf in (\ref{eqn:DivergenceAlgoSimp2}), we have:
\[\frac{\partial}{\partial \theta}\hat{D}(p_{\lambda^{k+1},\theta^{k+1}},p_T) + \frac{\partial}{\partial \theta} D_{\psi}((\lambda^{k+1},\theta^{k+1}),\phi^k) = 0\quad \forall k\]
Using the continuity of the derivatives (Assumptions A1 and AC), we may pass to the limit inside the gradients:
\[\frac{\partial}{\partial \theta}\hat{D}(p_{\tilde{\lambda},\tilde{\theta}},p_T) + \frac{\partial}{\partial \theta} D_{\psi}((\tilde{\lambda},\tilde{\theta}),\bar{\phi}) = 0\quad \forall k\]
As in the proof of Proposition \ref{prop:PhiDiffConverge}, all terms in the gradient of $D_{\psi}$ depend on $\psi'\left(\frac{h_i(x|\tilde{\lambda}, \tilde{\theta})}{h_i(x|\bar{\phi})}\right)$ which is zero by virtue of (\ref{eqn:ProportionsEqualityAll}). Hence $\frac{\partial}{\partial \theta}\hat{D}(p_{\tilde{\lambda},\tilde{\theta}},p_T) = 0$.\\
We prove now that $\frac{\partial}{\partial \lambda}\hat{D}(p_{\tilde{\lambda},\tilde{\theta}},p_T) = 0$. This is basically ensured by recurrence (\ref{eqn:DivergenceAlgoSimp1}), identities (\ref{eqn:ProportionsEqualityAll}), assumptions A1-AC and the fact that $\psi'(1)=0$. Indeed, using recurrence (\ref{eqn:DivergenceAlgoSimp1}), $\lambda^{k+1}$ is an optimum so that the gradient of the objective function is zero:
\[\frac{\partial}{\partial\lambda} \hat{D}(p_{\lambda^{k+1},\theta^k},p_T) + \frac{\partial}{\partial\lambda} D_{\psi}((\lambda^{k+1},\theta^k),\lambda^k,\theta^k) = 0, \quad \forall k\]
Since $\|\theta^{k+1}-\theta^k\|\rightarrow 0$, then $\bar{\theta} = \tilde{\theta}$. By passing to the limit in the previous identity and using the continuity of the derivatives, we have:
\[\frac{\partial}{\partial\lambda} \hat{D}(p_{\tilde{\lambda},\bar{\theta}},p_T) +  \frac{\partial}{\partial\lambda} D_{\psi}((\tilde{\lambda},\tilde{\theta}),\bar{\lambda},\bar{\theta}) = 0\]
Since the derivative of $D_{\psi}$ is a sum of terms which depend all on $\psi'(\frac{h_i(x|\tilde{\lambda},\bar{\theta})}{h_i(|\bar{\lambda},\bar{\theta})})$, and using identities (\ref{eqn:ProportionsEqualityAll}), we conclude that $\psi'(\frac{h_i(|\tilde{\lambda},\bar{\theta})}{h_i(|\bar{\lambda},\bar{\theta})})=\psi'(1)=0$ and $\frac{\partial}{\partial\lambda} D_{\psi}((\tilde{\lambda},\bar{\theta}),\bar{\lambda},\bar{\theta}) = 0$. Finally, $\bar{\theta}=\tilde{\theta}$ implies that $\frac{\partial}{\partial\lambda} \hat{D}(p_{\tilde{\lambda},\hat{\theta}},p_T) = 0$.\\
We have proved that $\frac{\partial}{\partial \lambda}\hat{D}(p_{\tilde{\lambda},\tilde{\theta}},p_T) = 0$ and $\frac{\partial}{\partial \theta}\hat{D}(p_{\tilde{\lambda},\tilde{\theta}},p_T) = 0$, so the gradient is zero and the stated result is proved.
\end{IEEEproof}

\begin{remark}
The previous proposition demands a condition on the distance between two consecutive members of the sequence $(\theta^k)_k$ which is \emph{a priori} weaker than the same condition on the whole sequence $\phi^k=(\lambda^k,\theta^k)$. Still, as the regularization term $D_{\psi}$ does not verify the identifiability condition A3, it stays an open problem for a further work. It is interesting to notice that condition $\|\theta^{k+1}-\theta^k\|\rightarrow 0$ can be replaced by $\|\lambda^{k+1}-\lambda^k\|\rightarrow 0$, but we then need to change the order of steps (\ref{eqn:DivergenceAlgoSimp1}) and (\ref{eqn:DivergenceAlgoSimp2}). A condition over the proportions seems to be \emph{simpler}.
\end{remark}
\begin{remark}
We can define an algorithm which converges to a global infimum of the estimated $\varphi-$divergence. The idea is very simple. We need to multiply the proximal term by a sequence $(\beta_k)_k$ of positive numbers which decreases to zero, for example $\beta_k=1/k$. The justification of such variant can be deduced from Theorem 3.2.4 in \cite{ChretienHeroProxGener}. The problem with this approach is that it depends heavily on the fact that the supremum on each step of the algorithm is calculated exactly. This does not happen in general unless function $\hat{D}(p_{\phi},p_T) + \beta_k D_{\psi}(\phi,\phi^k)$ is strictly convex. Although in our approach, we use similar assumption to prove the consecutive decreasing of $\hat{D}(p_{\phi},p_T)$, we can replace the infimum calculus in (\ref{eqn:DivergenceAlgo}) by two things. We require at each step that we find a local infimum of $\hat{D}(p_{\phi},p_T) + D_{\psi}(\phi,\phi^k)$ whose evaluation with $\phi\mapsto\hat{D}(p_{\phi},p_T)$ is less than the previous term of the sequence $\phi^k$. If we can no longer find any local maxima verifying the claim, the procedure stops with $\phi^{k+1}=\phi^k$. This ensures the availability of all proofs presented in this paper with no further changes.
\end{remark}

\section{Examples}\label{sec:Examples}
\subsection{Two-component Gaussian mixture}\label{Example:GaussMix}
We suppose that the model $(p_{\phi})_{\phi\in\Phi}$ is a mixture of two Gaussian densities, and suppose that we are only interested in estimating the means $\mu=(\mu_1,\mu_2)\in\mathbb{R}^2$ and the proportions $\lambda = (\lambda_1,\lambda_2)\in[\eta,1-\eta]^2$. The use of $\eta$ is to avoid cancellation of any of the two components and to keep the hypothesis about the conditional densities $h_i$ true, i.e. $h_i(x|\phi)>0$ for $x=1,2$. We also suppose to simplify the calculus that the components variances are reduced ($\sigma_i = 1$). The model takes the form:
\begin{equation}
p_{\lambda,\mu}(x) = \frac{\lambda}{\sqrt{2\pi}} e^{-\frac{1}{2}(x-\mu_1)^2} + \frac{1-\lambda}{\sqrt{2\pi}} e^{-\frac{1}{2}(x-\mu_2)^2},
\label{eqn:GaussMixModel}
\end{equation}
where $\Phi = [\eta,1-\eta]^s\times\mathbb{R}^s$. Here $\phi=(\lambda,\mu_1,\mu_2)$. The distance-like function $D_{\psi}$ is defined by:
\[D_{\psi}(\phi,\phi^k) = \sum_{i=1}^n{\psi\left(\frac{h_i(1|\phi)}{h_i(1|\phi^k)}\right)h_i(1|\phi^k)} + \sum_{i=1}^n{\psi\left(\frac{h_i(2|\phi)}{h_i(2|\phi^k)}\right)h_i(2|\phi^k)},\]
where:
\[h_i(1|\phi) = \frac{\lambda e^{-\frac{1}{2}(y_i-\mu_1)^2}}{\lambda e^{-\frac{1}{2}(y_i-\mu_1)^2} + (1-\lambda) e^{-\frac{1}{2}(y_i-\mu_2)^2}}, \quad h_i(2|\phi) = 1-h_i(1|\phi).\]
It is clear that functions $h_i$ are of class $\mathcal{C}^1$ on (int($\Phi$)), and as a consequence, $D_{\psi}$ is also of class $\mathcal{C}^1$ on (int($\Phi$)).\\
\textbf{If we use the MDPD (\ref{eqn:MDPDdef})}, then function $\phi\mapsto\hat{D}(p_{\phi},p_T)$ is clearly continuously differentiable by Lebesgue theorems. Recall that $\hat{D}_a(p_{\phi},p_T)$ is given by $\int{p_{\lambda,\mu}^{1+a}(y)dy} - \frac{a+1}{a}\frac{1}{n}\sum{p_{\lambda,\mu}^a(y_i)}$, since we dropped the supplementary term $\frac{1}{a}\int{p_T^{1+a}(y)dy}$ from (\ref{eqn:DPD}) because it does not depend on the parameters. Notice that for any $\mu=(\mu_1,\mu_2)$ such that $\|\mu\|<M$, $p_{\phi}^{1+a}(y)\leq ce^{-(1+a)y^2}$ for some positive constant $c$ which depends on $M$ and $a$. Thus, assumptions A1 and AC are verified. In order to prove that $\Phi^0$ is compact, we prove that it is closed and bounded in the complete space $[\eta,1-\eta]\times\mathbb{R}^2$.  Closedness is an immediate result of the continuity of the estimated divergence. Indeed,
\begin{eqnarray*}
\Phi^0 & = & \left\{\phi\in\Phi, \hat{D}_a(p_{\phi},p_T)\leq \hat{D}_(p_{\phi^0},p_T)\right\} \\
			 & = & \hat{D}_a(p_{\phi},p_T)^{-1}\left((-\infty, \hat{D}_{\varphi}(p_{\phi^0},p_T)]\right).
\end{eqnarray*}
In order to ensure boundedness of $\Phi^0$, we need to choose carefully the initial point $(\lambda^0,\mu^0)$ of the algorithm. Since $\lambda$ is bounded by 0 and 1, we only need to verify the boundedness of the means. If both means $\mu_1$ and $\mu_2$ go to $\pm\infty$, then $\hat{D}_a(p_{\phi,p_T})\rightarrow 0$. Besides, if either of the means go to $\pm\infty$, then the corresponding component vanishes. Thus if we choose $(\lambda^0,\mu^0)$ such that:
\begin{equation}
\hat{D}_a(p_{(\lambda^0,\mu_1^0,\mu_2^0)},p_T) < \min\left(0,\inf_{\lambda\in[\eta,1-\eta],\mu_1\in\mathbb{R}}\hat{D}_a(p_{(\lambda,\mu_1,\infty)},p_T)\right),
\label{eqn:CondGaussMixDPD}
\end{equation}
then by definition of $\Phi^0$, any point of it must have a corresponding value of $\hat{D}_a(p_{\phi,p_T})$ less than its values at the extremities, i.e. when either of both means goes to infinity. Thus, under condition (\ref{eqn:CondGaussMixDPD}), $\Phi^0$ is bounded. Now that assumption A2 is also fulfilled, we arrive to the following conclusion.
\begin{conclusion}
\label{conc:conclusion0}
Using Propositions \ref{prop:DecreaseDphi} and \ref{prop:NewRes} and under condition (\ref{eqn:CondGaussMixDPD}), the sequence $(\hat{D}_a(p_{\phi^k},p_T))_k$ converges and there exists a subsequence $(\phi^{N(k)})$ which converges to a stationary point of the estimated divergence. Moreover, every limit point of the sequence $(\phi^k)_k$ is a stationary point of the estimated divergence. 
\end{conclusion}

\textbf{If we are using the dual estimator of the $\varphi-$divergence given by (\ref{eqn:DivergenceDef})}. This was discussed in \cite{DiaaBroniaProximalEntropy}. We cite only the final conclusion for the sequence $(\phi^k)_k$ defined by any of the proximal-point algorithms (\ref{eqn:DivergenceAlgo}) or (\ref{eqn:DivergenceAlgoSimp1},\ref{eqn:DivergenceAlgoSimp2}).
\begin{conclusion}
\label{conc:conclusion1}
Using Propositions \ref{prop:DecreaseDphi} and \ref{prop:NewRes}, if $\Phi=[\eta,1-\eta]\times [\mu_{\min},\mu_{\max}]^2$, the sequence $(\hat{D}_{\varphi}(p_{\phi^k},p_T))_k$ defined through formula (\ref{eqn:DivergenceDef}) converges and there exists a subsequence $(\phi^{N(k)})$ which converges to a stationary point of the estimated divergence. Moreover, every limit point of the sequence $(\phi^k)_k$ is a stationary point of the estimated divergence. 
\end{conclusion}
\textbf{If we are using the kernel-based dual estimator given by (\ref{eqn:NewDualForm})} with a Gaussian kernel density estimator, then if the we initialize any of the proximal-point algorithms (\ref{eqn:DivergenceAlgo}) or (\ref{eqn:DivergenceAlgoSimp1},\ref{eqn:DivergenceAlgoSimp2}) with $\phi^0$ verifying:
\begin{eqnarray}
\hat{D}_{\varphi}(p_{\phi},p_T) & < & \min\left(\frac{1}{\gamma(\gamma-1)}, \inf_{\lambda,\mu}\hat{D}_{\varphi}(p_{(\lambda,\infty,\mu)},p_T)\right)\qquad \text{ if } \gamma \in(0,\infty)\setminus \{1\};\label{eqn:CondGaussMixNewDual1} \\
\hat{D}_{\varphi}(p_{\phi},p_T) & < & \inf_{\lambda,\mu}\hat{D}_{\varphi}(p_{(\lambda,\infty,\mu)},p_T)\qquad \text{ if } \gamma <0,
\label{eqn:CondGaussMixNewDual2}
\end{eqnarray}
we have the following conclusion (see \cite{DiaaBroniaProximalEntropy})
\begin{conclusion}
\label{conc:conclusion2}
Using Propositions \ref{prop:DecreaseDphi} and \ref{prop:NewRes}, under condition (\ref{eqn:CondGaussMixNewDual1}, \ref{eqn:CondGaussMixNewDual2}) the sequence $(\hat{D}_{\varphi}(p_{\phi^k},p_T))_k$ defined through formula (\ref{eqn:NewDualForm}) converges and there exists a subsequence $(\phi^{N(k)})$ which converges to a stationary point of the estimated divergence. Moreover, every limit point of the sequence $(\phi^k)_k$ is a stationary point of the estimated divergence. 
\end{conclusion}
\textbf{In the case of the likelihood $\varphi(t)=-\log(t)+t-1$}, then if we initialize any of the proximal-point algorithms (\ref{eqn:DivergenceAlgo}) or (\ref{eqn:DivergenceAlgoSimp1},\ref{eqn:DivergenceAlgoSimp2}) with $\phi^0$ verifying:
\begin{equation}
J(\phi^0)>\max\left[J\left(0,\infty,\frac{1}{n}\sum_{i=1}^n{y_i}\right),\; J\left(1,\frac{1}{n}\sum_{i=1}^n{y_i},\infty\right)\right]
\label{eqn:TwoGaussMixCond}
\end{equation}
then we reach the following conclusion (see\cite{AlMohamad2015})
\begin{conclusion}
\label{conc:conclusion3}
Using Propositions \ref{prop:DecreaseDphi} and \ref{prop:NewRes}, under condition (\ref{eqn:TwoGaussMixCond}) the sequence $(J(\phi^k))_k$ converges and there exists a subsequence $(\phi^{N(k)})$ which converges to a stationary point of the likelihood function. Moreover, every limit point of the sequence $(\phi^k)_k$ is a stationary point of the likelihood. 
\end{conclusion}
\textbf{Assumption A3 is not fulfilled} (this part applies for all aforementioned situations). We study  the equation $D_{\psi}(\phi|\phi')=0$ for $\phi\neq\phi'$. By definition of $D_{\psi}$, it is given by a sum of nonnegative terms, which implies that all terms need to be equal to zero. The following lines are equivalent $\forall i \in \{1,\cdots,n\}$:
\begin{eqnarray*}
h_i(0|\lambda,\mu_1,\mu_2) & = & h_i(0|\lambda',\mu'_1,\mu'_2); \\
\frac{\lambda e^{-\frac{1}{2}(y_i-\mu_1)^2}}{\lambda e^{-\frac{1}{2}(y_i-\mu_1)^2} + (1-\lambda) e^{-\frac{1}{2}(y_i-\mu_2)^2}} & = & \frac{\lambda' e^{-\frac{1}{2}(y_i-\mu'_1)^2}}{\lambda' e^{-\frac{1}{2}(y_i-\mu'_1)^2} + (1-\lambda') e^{-\frac{1}{2}(y_i-\mu'_2)^2}}; \\
\log\left(\frac{1-\lambda}{\lambda}\right) - \frac{1}{2}(y_i-\mu_2)^2 + \frac{1}{2}(y_i-\mu_1)^2 & = & \log\left(\frac{1-\lambda'}{\lambda'}\right) - \frac{1}{2}(y_i-\mu'_2)^2 + \frac{1}{2}(y_i-\mu'_1)^2.
\end{eqnarray*}
Looking at this set of $n$ equations as an equality of two polynomials on $y$ of degree 1 at $n$ points\footnote{The second order terms vanish from both sides of the each equation.}, we deduce that as we dispose of two distinct observations, say, $y_1$ and $y_2$, the two polynomials need to have the same coefficients. Thus the set of $n$ equations is equivalent to the following two equations:
\begin{equation}
\left\{\begin{array}{ccc}\mu_1-\mu_2 & = & \mu'_1-\mu'_2 \\
		\log\left(\frac{1-\lambda}{\lambda}\right) + \frac{1}{2}\mu_1^2 - \frac{1}{2}\mu_2^2 & = & \log\left(\frac{1-\lambda'}{\lambda'}\right) + \frac{1}{2}{\mu'_1}^2 - \frac{1}{2}{\mu'_2}^2	
		\end{array}\right.
\label{eqn:EqSysGaussMix}
\end{equation}
These two equations with three variables have an infinite number of solutions. Take for example $\mu_1 = 0,\mu_2=1,\lambda=\frac{2}{3},\mu'_1=\frac{1}{2}, \mu'_2=\frac{3}{2},\lambda'=\frac{1}{2}$. This entails that, for any $\phi\in\Phi$, there exists an infinite number of elements in $\Phi$ for which the value of $D_{\psi}$ between $\phi$ and any one of them is equal to zero. This proves that assumption A3 is not fulfilled in the Gaussian mixture.\\
\begin{remark} 
The previous conclusion can be extended to any two-component mixture of exponential families having the form:
\[p_{\phi}(y) = \lambda e^{\sum_{i=1}^{m_1}{\theta_{1,i}y^{i}} - F(\theta_1)} + (1-\lambda)e^{\sum_{i=1}^{m_2}{\theta_{2,i}y^{i}} - F(\theta_2)}.\]
One may write the corresponding $n$ equations. The polynomial of $y_i$ has a degree of at most $\max(m_1,m_2)$. Thus, if one disposes of $\max(m_1,m_2)+1$ distinct observations, the two polynomials will have the same set of coefficients. Finally, if $(\theta_1,\theta_2)\in\mathbb{R}^{d-1}$ with $d>\max(m_1,m_2)$, then assumption A3 does not hold.
\end{remark}
This conclusion holds for both algorithms (\ref{eqn:DivergenceAlgo}) or (\ref{eqn:DivergenceAlgoSimp1},\ref{eqn:DivergenceAlgoSimp2}). Unfortunately, we have no information about the difference between consecutive terms $\|\phi^{k+1}-\phi^k\|$ except for the case of $\psi(t) = \varphi(t)=-\log(t)+t-1$ which corresponds to the classical EM recurrence:
\[\lambda^{k+1} = \frac{1}{n}\sum_{i=1}^n{h_i(0|\phi^k)},\quad \mu_1^{k+1} = \frac{\sum_{i=1}^n{y_ih_i(0|\phi^k)}}{\sum_{i=1}^n{h_i(0|\phi^k)}}\quad \mu_1^{k+1} = \frac{\sum_{i=1}^n{y_ih_i(1|\phi^k)}}{\sum_{i=1}^n{h_i(1|\phi^k)}}.\]
In such case, \cite{Tseng} has shown that we can prove directly that $\phi^{k+1}-\phi^k$ converges to 0 without the use of Proposition \ref{prop:PhiDiffConverge}.

\subsection{Two-component Weibull mixture}\label{subsec:WeibullMixEx}
Let $p_{\phi}$ be a two-component Weibull mixture:
\begin{equation}
p_{\phi}(x) = 2\lambda\phi_1 (2x)^{\phi_1-1} e^{-(2x)^{\phi_1}}+(1-\lambda)\frac{\phi_2}{2}\left(\frac{x}{2}\right)^{\phi_2-1} e^{-\left(\frac{x}{2}\right)^{\phi_2}}, \qquad \phi=(\lambda,\phi_1,\phi_2).
\label{eqn:WeibullMixture}
\end{equation}
We have $\Phi = (0,1)\times\mathbb{R}_+^*\times\mathbb{R}_+^*$. Similarly to the Gaussian example, we will study convergence properties in light of our theoretical approach. We will only be interested in power divergences ddefined through the Cressie-Read class of functions $\varphi=\varphi_{\gamma}$ given by (\ref{eqn:CressieReadPhi}).\\
The weight functions $h_i$ are given by:
\[h_i(1|\phi) = \frac{2\lambda\phi_1 (2x)^{\phi_1-1} e^{-(2x)^{\phi_1}}}{2\lambda\phi_1 (2x)^{\phi_1-1} e^{-(2x)^{\phi_1}}+(1-\lambda)\frac{\phi_2}{2}\left(\frac{x}{2}\right)^{\phi_2-1} e^{-\left(\frac{x}{2}\right)^{\phi_2}}},\quad h_i(2|\phi)=1-h_i(1|\phi).\]
It is clear the functions $h_i$ are of class $\mathcal{C}^1(\text{int}(\Phi))$ and so does $\phi\mapsto D_{\psi}(\phi,\phi')$ for any $\phi'\in\Phi$.\\
\textbf{If we use the MDPD (\ref{eqn:MDPDdef})}, the continuity and differentiability of the estimated divergence $\hat{D}_a$ can be treated similarly to the Gaussian example. The proof of compactness of $\Phi^0$ is also similar. We identify a condition on the initialization of the algorithm in order to make $\Phi^0$ bounded.
\[\hat{D}_a(p_{\lambda,\phi},p_T) < \min\left(0,\inf_{\phi_1>0,\lambda\in[\eta,1-\eta]}\hat{D}_a(p_{(\lambda,\phi_1,\infty)},p_T)\right).\]
A conclusion similar to Conclusion \ref{conc:conclusion0} can be stated here.\\
\textbf{If we are using the dual estimator defined by (\ref{eqn:DivergenceDef})}, then continuity can be treated similarly to the case of the Gaussian example. Here, however, the continuity and differentiability of the optimized function $f(\alpha,\phi)$, where $\hat{D}_{\varphi}(p_{\phi},p_T) = \sup_{\alpha}f(\alpha,\phi)$, are more technical. We list the following three results without any proof, because it suffices to study the integral term in the formula. Suppose, without loss of generality, that $\phi_1<\phi_2$ and $\alpha_1<\alpha_2$.
\begin{enumerate}
\item For $\gamma>1$, which includes the Pearson's $\chi^2$ case, the dual representation is \emph{not} well defined since $\sup_{\alpha}f(\alpha,\phi)=\infty$;
\item For $\gamma\in(0,1)$, function $f(\alpha,\phi)$ is continuous. 
\item For $\gamma<0$, function $f(\alpha,\phi)$ is continuous and well defined for $\phi_1<\frac{\gamma-1}{\gamma}\alpha_1$ and $\alpha_2\geq \phi_2$. Otherwise $f(\alpha,\phi)=-\infty$, but the supremum $\sup_{\alpha}f(\alpha,\phi)$ is still well defined.
\end{enumerate}
In both cases 2 and 3, differentiability of function $f(\alpha,\phi)$ holds only on a subset of $\Phi\times\Phi$ which cannot be written as $A\times B$, and thus the theoretical approaches presented in Section \ref{sec:AnalyticalDiscuss} are not suitable. In order to end this part, we emphasize the fact that, similarly to the Gaussian example, even continuity of the estimated divergence $\hat{D}_{\varphi}(p_{\phi},p_T)$ with respect to $\phi$ cannot be directly using the theoretical approaches presented in paragraph (\ref{sec:AnalyticalDiscuss}) unless we suppose that $\Phi$ is compact. Indeed, if $\Phi$ is compact, then using Theorem 1.17 from \cite{Rockafellar}, continuity of the estimated divergence is a direct result. Differentiability of the estimated divergence is far more difficult and needs more investigations on the form of the estimated divergence and the model used.\\
Similar conclusion to Conclusion \ref{conc:conclusion1} can be stated here with no changes except for the fact that assumption AC is not fulfilled. This entails that our conclusion will be about the subgradient of the estimated divergence.\\
\textbf{If we are using the kernel-based dual estimator given by (\ref{eqn:NewDualForm})} with a Gaussian kernel density estimator, then things are a lot simplified. We need only to treat the integral term. From an analytic point of view, the study of continuity depends on the kernel used; more specifically its tail behavior. If we take a Gaussian kernel, then we have:
\begin{itemize}
\item[$\bullet$] For $\gamma>1$, it is necessary that $\min(\phi_1,\phi_2)>2$, otherwise the estimated divergence is infinity. Thus, it is necessary for either of the true values of the shapes to be inferior to 2 in order for the estimation to be valid;
\item[$\bullet$] For $\gamma\in(0,1)$, then the estimated divergence is $\mathcal{C}^1(\text{int}(\Phi))$;
\item[$\bullet$] For $\gamma<0$, it is necessary that $\min(\phi_1,\phi_2)<1-\frac{1}{\gamma}$ and $\max(\phi_1,\phi_2)<2$. If these conditions do not hold, then the estimated divergence is minimized at $-\infty$ at any vector of parameters which does not verify the previous condition.
\end{itemize}
In the first case, if we use a heavier-tailed kernel such as the Cauchy Kernel, the estimated divergence becomes $\mathcal{C}^1(\text{int}(\Phi))$. In the third case, if we use a compact-supported kernel such as the Epanechnikov's kernel, the condition is reduced to only $\min(\phi_1,\phi_2)<1-\frac{1}{\gamma}$.\\
Similar conditions to (\ref{eqn:CondGaussMixNewDual1},\ref{eqn:CondGaussMixNewDual2}) can be obtained and we have the same conclusion as Conclusion \ref{conc:conclusion2}. \\
\textbf{In the case of the Likelihood} $\varphi(t)=-\log(t)+t-1$, we illustrate the convergence of the EM algorithm through our theoretical approach. Assumptions A1 and AC are clearly verified since both the log-likelihood and the proximal term are sums of continuously differentiable functions, and integrals do not intervene here. The set $\Phi^0$ is given by: 
\begin{eqnarray*}
\Phi^0 & = & \left\{\phi\in\Phi, J(\phi)\geq J(\phi^0)\right\} \\
       & = & J^{-1}\left([J(\phi^0),\infty)\right) \\
			 & = & \left\{\phi\in\Phi, L(\phi)\geq L(\phi^0)\right\}
\end{eqnarray*}
where $L(\phi)$ is the likelihood of the model, and $J(\phi) = \log(L(\phi))$ is the log-likelihood function. We will show that under similar conditions to the Gaussian mixture, the set $\Phi^0$ is compact. \\
\textbf{Closedness of $\Phi^0$}. Since the shape parameter is supposed to be positive, continuity of the log-likelihood would imply only that $\Phi^0$ is closed in $[0,1]\times\mathbb{R_+^*}\times\mathbb{R_+^*}$, a space which is not closed and hence is not complete. We therefore, propose to extend the definition of shape parameter on 0. From a statistical point of view, this extension is not reasonable since the density function of Weibull distribution with a shape parameter equal to 0 is the zero function which is not a probability density. Besides, identifiability problems would appear for a mixture model. Nevertheless, our need is only for analytical purpose. We will add suitable conditions on $\phi^0$ in order to avoid such subtlety keeping in hand the closedness property.\\
We suppose now that the shape parameter can have values in $\mathbb{R_+}$. The set $\Phi^0$ is now the inverse image of $[L(\phi^0),\infty)$ by the likelihood function\footnote{We do not use this time the log-likelihood function since it is not defined when both shape parameters are zero.} which is continuous on $[0,1]\times\mathbb{R_+}\times\mathbb{R_+}$. Hence, it is closed in the space $[0,1]\times\mathbb{R_+}\times\mathbb{R_+}$ provided the euclidean norm which is complete. It suffices then to prove that $\Phi^0$ is bounded.\\
\textbf{Boundedness of $\Phi^0$.} We will make similar arguments to the case of the Gaussian mixture example. We need to calculate the limit at infinity when the shape parameter of either of the two components tends to infinity. If both $\phi_1$ and $\phi_2$ goes to infinity, the log-likelihood tends to $-\infty$. Hence any choice of a finite $\phi^0$ can avoid this case. Suppose now that $\phi_1$ goes to infinity whereas $\phi_2$ stays bounded. The corresponding limit of the log-likelihood functions is given by:
\[J(\lambda,\infty,\phi_2) = \sum_{i=1}^n{\log\left((1-\lambda)\frac{\phi_2}{2}\left(\frac{y_i}{2}\right)^{\phi_2-1} e^{-\left(\frac{y_i}{2}\right)^{\phi_2}}\right)}\]
if there is no observation $y_i$ equal to $\frac{1}{2}$. In fact, if there is $y_i=\frac{1}{2}$, the limit is $+\infty$ and the set $\Phi^0$ cannot be bounded. However, it is improbable to get such an observation since the probability of getting an observation equal to $\frac{1}{2}$ is zero. The case when $\phi_2$ goes to infinity whereas $\phi_1$ stays bounded is treated similarly. \\
To avoid the two previous scenarios, one should choose the initial point of the algorithm $\phi^0$ in a way that it verifies:
\begin{equation}
J(\phi^0)> \max\left(\sup_{\lambda,\phi_2} J(\lambda,\infty,\phi_2), \sup_{\lambda,\phi_1} J(\lambda,\phi_1,\infty)\right).
\label{eqn:ConditionWeibullMix}
\end{equation}
Since all vectors of $\Phi^0$ have a log-likelihood value greater than $J(\phi^0)$, the previous choice permits the set $\Phi^0$ to avoid non-finite values of $\phi$. Thus it becomes bounded whenever $\phi_0$ is chosen according to condition (\ref{eqn:ConditionWeibullMix}). Finally, the calculus of both terms $\sup_{\lambda,\phi_1} J(\lambda,\phi_1,\infty)$ and $\sup_{\lambda,\phi_2} J(\lambda,\phi_2,\infty)$ is not feasible but numerically. They, however, can be simplified a little. One can notice by writing these terms without the logarithm (as a product), the term which has $\lambda$ is maximized when it is equal to 1. The remaining of the calculus is a maximization of the likelihood function of a Weibull model\footnote{In a Weibull model, the calculus of the MLE cannot be done but numerically when the parameter of interest is the shape parameter.}.\\
We conclude that the set $\Phi^0$ is compact under condition (\ref{eqn:ConditionWeibullMix}). Finally, it is important to notice that condition (\ref{eqn:ConditionWeibullMix}) permits also to avoid the border values which corresponds to $\phi_1=0$ or $\phi_2=0$. Indeed, when either of the shape parameters is zero, the corresponding component vanishes and the corresponding log-likelihood value is less than the upper bound in condition (\ref{eqn:ConditionWeibullMix}). The same conclusion as Conclusion \ref{conc:conclusion3} can be stated here for the Weibull mixture model.\\
Notice that the verification of assumption A3 is a hard task here because it results in a set of $n$ nonlinear equations in $y_i$ and cannot be treated in a similar way to the Gaussian mixture.
\subsection{Pearson's \texorpdfstring{$\chi^2$}{chi square} algorithm for a Cauchy model}\label{Example:Cauchy}
Let $\{(x_i,y_i),i=0,\cdots,n\}$ be an n-sample drawn from the joint probability law defined by the density function:
\[f(x,y|a,x_0) = \frac{a(y-x_0)^2e^x}{\pi\left(a^2+(y-x_0)^2e^x\right)^2}, \quad x\in[0,\infty), y\in\mathbb{R}\]
where $a\in[\varepsilon,\infty)$, with $\varepsilon>0$, denotes a scale parameter and $x_0\in\mathbb{R}$ denotes a location parameter. We define an exponential probability law with parameter $\frac{1}{2}$ on the labels. It is given by the density function:
\[q(x)=\frac{1}{2}e^{-x/2}.\]
Now, the model defined on the observed data becomes a Cauchy model with two parameters:
\[p_{(a,x_0)}(y) = \int_{0}^{\infty}{f(x,y|a,x_0)dx} = \frac{a}{\pi(a^2+(y-x_0)^2)},\quad a\geq\varepsilon>0, x_0\in\mathbb{R}.\]
The goal of this example is to show how we prove assumptions A1-3 and AC in order to explore the convergence properties of the sequence $\phi^k$ generated by either of the algorithms (\ref{eqn:DivergenceAlgo}) and (\ref{eqn:DivergenceAlgoSimp1},\ref{eqn:DivergenceAlgoSimp2}). We also discuss the analytical properties of the dual representation of the divergence.\\
In this example, we only focus on the dual representation of the divergence given by (\ref{eqn:DivergenceDef}) because the resulting MD$\varphi$DE is robust against outliers (so does the MLE). Thus there is no need to use a robust estimator such as the kernel-based MD$\varphi$DE which needs a choice of a suitable kernel and window.
\subsubsection{Cauchy model with zero location}\label{Example:CauchyZeroLoc}
\begin{figure}[ht]
\centering
\includegraphics[scale = 0.35]{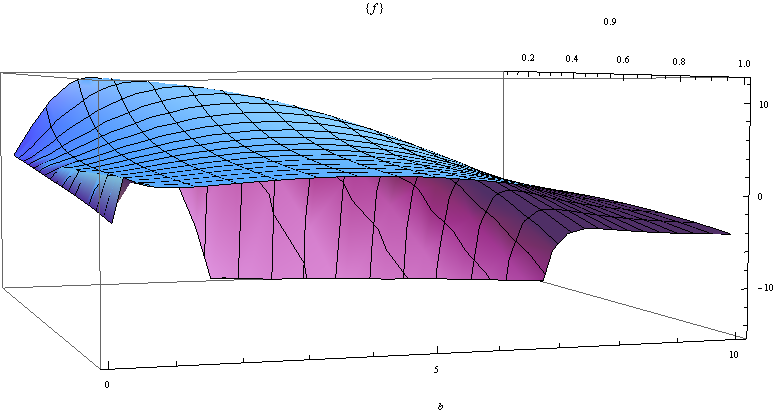}
\caption{A 3D plot of function $f(a,b)$ for a 10-sample of the standard Cauchy distribution.}
\label{fig:CauchyExample3D}
\end{figure}

 We suppose here that $x_0=0$, and we are only interested in estimating the scale parameter $a$. The Pearson's $\chi^2$ divergence is given by:
\[D(p_{a},p_{a^*}) = \frac{1}{2}\int{\left[\frac{p_{a}(y)}{p_{a^*}}-1\right]^2p_{a^*}(y)dy}.\]
Let's rewrite the dual representation of the Chi square divergence:
\[\hat{D}(p_{a},p_{a^*}) = \sup_{b\geq\varepsilon}\left\{\int_{\mathbb{R}}{\frac{p_b^2(x)}{p_a(x)}dx} - \frac{1}{2n}\sum_{i=1}^n{\frac{p_b^2(y_i)}{p_a^2(y_i)}}\right\} - \frac{1}{2}.\]
A simple calculus shows:
\[\int_{\mathbb{R}}{\frac{p_b^2(x)}{p_a(x)}dx} = \frac{(a^2+b^2)\pi}{2ab}.\]
This implies a simpler form for the dual representation of the divergence:
\begin{equation}
\hat{D}(p_{a},p_{a^*}) = \sup_{b\geq\varepsilon}\left\{\frac{(a^2+b^2)}{2ab} - \frac{1}{2n}\sum_{i=1}^n{\frac{a^2(b^2+y_i^2)^2}{b^2(a^2+y_i^2)^2}}\right\} - \frac{1}{2}.
\label{eqn:DivergenceCauchy}
\end{equation}
Let $f(a,b)$ denote the optimized function in the above formula. We calculate the first derivative with respect to $b$:
\[\frac{\partial f}{\partial b}(a,b) = -\frac{\pi a}{2b^2} + \frac{\pi}{2a} - \frac{1}{2n}\sum_{i=1}^n{\frac{a^2}{(a^2+y_i^2)^2}\left(2b-\frac{2y_i^4}{b^3}\right)}.\]
Notice that as $a\geq\varepsilon$ the term $\frac{\pi}{2a}$ stays bounded away from infinity uniformly. Therefore, it suffices then that $b$ exceeds a finite value $b_0$ in order that the derivative becomes negative. Hence, there exists $b_0$ such that $b\mapsto f(a,b)$ becomes decreasing independently of $a$. On the other hand $\forall a>0, \lim_{b\rightarrow\infty} f(a,b) = -\infty$. It results that all values of the function $b\mapsto f(a,b)$ for $b>b_0$ does not have any use in the calculus of the supremum in (\ref{eqn:DivergenceCauchy}), since, by the decreasing property if $b\mapsto f(a,b)$, they all should have values less than the value at $b_0$. We may now rewrite the dual representation of the Chi square divergence as :
\begin{equation}
\hat{D}(p_{a},p_{a^*}) = \sup_{b\in[\varepsilon,b_0]}\left\{\frac{(a^2+b^2)}{2ab} - \frac{1}{2n}\sum_{i=1}^n{\frac{a^2(b^2+y_i^2)^2}{b^2(a^2+y_i^2)}}\right\} - \frac{1}{2}.
\label{eqn:DualRepCauchyLocation}
\end{equation}
We have now two pieces of information about $f(a,b)$. The first is that it is level-bounded locally in $b$ uniformly in $a$ (see paragraph (\ref{para:LevelBoundFun})). The second is that we are exactly in the context of lower$-\mathcal{C}^1$ functions (\ref{para:LowerC1}). First of all, function $f$ is $\mathcal{C}^1([\varepsilon,\infty)\times[\varepsilon,\infty))$ function, so that part (a) of Theorem \ref{theo:levelbounded} is verified and the function $a\mapsto\hat{D}(p_{a},p_{a^*})$ is strictly continuous. To prove it is continuously differentiable, we need to prove that the set 
\[Y(a)=\bigcup_{b\in\argmax_{b'}f(a,b)}\left\{\frac{\partial f}{\partial a}(a,b)\right\}\]
contains but one element. From a theoretic point of view, two possible methods are available: Prove that either there is a unique maximum for a fixed $a$, or that the derivative with respect to $a$ at all maxima \emph{does not depend} on $a$ (they have the same value). In our example, function $b\mapsto f(a,b)$ is not concave. We may also plot it using any mathematical tool provided that we already have the data set. We tried out a simple example and generated a 10-sample of the standard Cauchy distribution ($a=1$), see table (\ref{tab:CauchyDerivEx}). We used Mathematica to draw a 3D figure of function $f$, see figure (\ref{fig:CauchyExample3D}).
\begin{table}[h]
\centering
\begin{tabular}{|c|c|c|c|c|c|c|c|c|c|c|}
$y_i$ & 0.534 & -18.197 & 0.726 & -0.439 & -1.945 & 0.0119 & 12.376 & -0.953 & 0.698 & 0.818\\
\hline
\end{tabular}
\caption{A 10-sample Cauchy dataset.}
\label{tab:CauchyDerivEx}
\end{table}

\begin{figure}[ht]
\centering
\includegraphics[scale = 0.2]{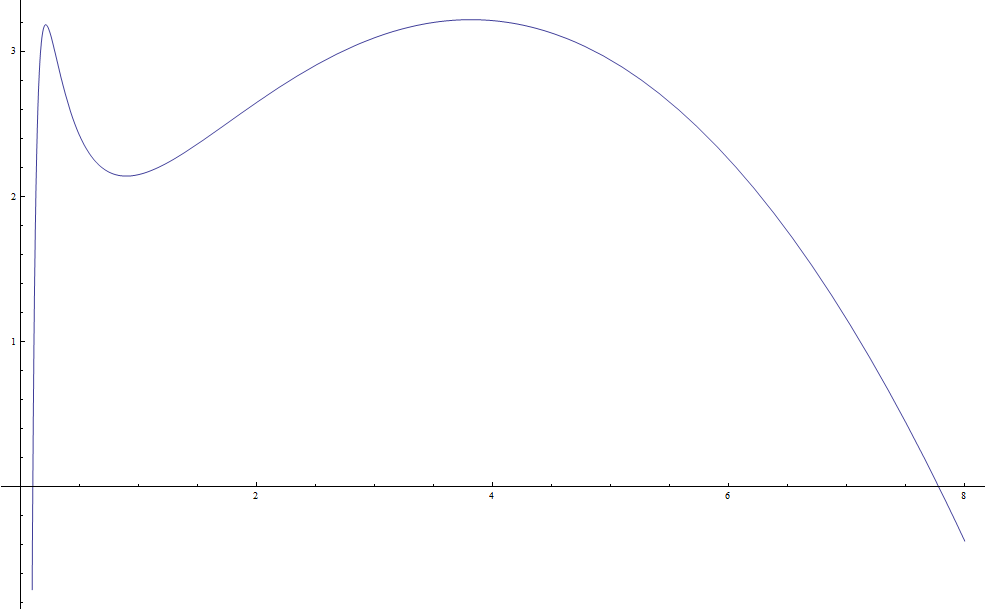}
\caption{A 2D plot of function $f(0.9,b)$ for a 10-sample of the standard Cauchy distribution.}
\label{fig:CauchyExample2D}
\end{figure}

It is clear that for a fixed $a$, the function $b\mapsto f(a,b)$ has two maxima which may both be global maxima. For example for $a=0.9$, one gets figure (\ref{fig:CauchyExample2D}). It is clearer now that conditions of Theorem \ref{theo:levelbounded} are not fulfilled, and we cannot prove that function $\hat{D}(p_{a},p_{a^*})$ is continuously differentiable every where.\\
\noindent It is however not the end of the road. We still have the results presented in paragraph (\ref{para:LowerC1}). Function $\hat{D}(p_{a},p_{a^*})$ is lower-$\mathcal{C}^1$. Therefore, it is strictly continuous and almost everywhere continuously differentiable. Hence, we may hope that the limit points of the sequence $(\phi^k)_k$ for algorithm (\ref{eqn:DivergenceAlgo}) are in the set of points where the dual representation of the Chi square divergence is $\mathcal{C}^1$, or be more reasonable and state any further result on the sequence in terms of the subgradient of $\hat{D}(p_{a},p_{a^*})$.\\
\paragraph{Compactness of $\Phi^0$.} We check when the set $\Phi^0=\{a|\hat{D}(p_{a},p_{a^*})\leq\hat{D}(p_{a_0},p_{a^*})\}$ is closed and bounded in $[\varepsilon,\infty)$ for an initial point $a_0$. \textbf{Closedness} is proved using continuity of $\hat{D}(p_{a},p_{a^*})$. Indeed,
\[\Phi^0 = \hat{D}^{-1}(p_{a},p_{a^*})\left((-\infty,\hat{D}(p_{a_0},p_{a^*})]\right).\]
\textbf{Boundedness} is proved by contradiction. Suppose that $\Phi^0$ is unbounded, then there exists a sequence $(a^l)_l$ of points of $\Phi^0$ which goes to infinity. Formula (\ref{eqn:DualRepCauchyLocation}) shows that $b$ stays in a bounded set during the calculus of the supremum. Hence the continuity of $\hat{D}(p_{a},p_{a^*})$ implies: 
\[\lim_{a\rightarrow\infty} \hat{D}(p_{a},p_{a^*}) = +\infty.\]
This shows that by choosing any finite $a_0$, the set $\Phi^0$ becomes bounded. Indeed, the relation defining $\Phi^0$ implies that $\forall l, \hat{D}(p_{a^l},p_{a^*})\leq\hat{D}(p_{a_0},p_{a^*})<\infty$, and a contradiction is reached by taking the limit of each part of this inequality. Hence $\Phi^0$ is closed and bounded in the space $[\varepsilon,\infty)$ which is complete provided with the euclidean distance. We conclude that $\Phi^0$ is compact\footnote{If we are to use a result which concerns the differentiability of $\hat{D}(p_{a},p_{a^*})$, one should consider the case when $\Phi^0$ shares a boundary with $\Phi$. A possible solution to avoid this is to consider an initial point $a^0$ such that $\hat{D}(p_{\varepsilon},p_{a^*})>\hat{D}(p_{a_0},p_{a^*})$. This expels the the boundary from the possible values of $\Phi^0$.}.\\
In this simple example, we only can use algorithm (\ref{eqn:DivergenceAlgo}) since there is only one parameter of interest. Proposition \ref{prop:NewRes} can be used to deduce convergence of any convergent subsequence to a generalized stationary point of $\hat{D}(p_{a},p_{a^*})$.\\
To deduce more results about the sequence $(a^k)_k$, we may try and verify assumption A3 using Lemma \ref{Lem:Tseng}. Let's write functions $h_i$.
\[h_i(x|a) = \frac{f(x,y_i|a)}{p_a(y_i)} = \frac{y_i^2e^x(a^2+y_i^2)}{(a^2+e^xy_i^2)^2}.\]
Clearly, for any $i\in\{1,\cdots,n\}$ and $a\geq\varepsilon$, function $x\mapsto h_i(x|a)$ is continuous. Let $a,b\geq\varepsilon$ such that $a\neq b$. Suppose that:
\[\forall i, \quad h_i(x|a)=h_i(x|b) \qquad \forall x\geq 0.\]
This entails that:
\[a^2b^4-a^4b2+(b^4-a^4)y_i^2+\left(a^2e^{2x}+2b^2e^x-b^2e^{2x}-2a^2e^x\right)y_i^4=0,\qquad i=1,\cdots,n.\]
This is a polynomial on $y_i$ of degree 4 which coincides with the zero polynomial on $n$ points. If there exists 5 distinct observations\footnote{If one uses the point $x=0$, the result follows directly without supposing the existence of distinct observations.}, then the two polynomials will have the same coefficients. Hence, we have $b^4-a^4=0$. This implies that $a=b$ since they are both positive real numbers. We conclude that $D_{\psi}(a,b)=0$ whenever $a=b$ which is equivalent to assumption A3. Proposition \ref{prop:PhiDiffConverge} can now be applied to deduce that sequence $(a^k)$ defined by (\ref{eqn:DivergenceAlgo}) (with $\phi^k$ replaced by $a^k$) is well defined and bounded. Furthermore, it verifies $a^{k+1}-a^k\rightarrow 0$, and the limit of any convergent subsequence is a generalized stationary point of $\hat{D}(p_{a},p_{a^*})$. The existence of such subsequence is guaranteed by the compactness of $\Phi^0$ and the fact that $\forall k, a^k\in\Phi^0$.

\section{Experimental results}\label{sec:Simulations}
We summarize the results of 100 experiments on $100$-samples (with and without outliers) from two-components Gaussian and Weibull mixtures by giving the average of the error committed with the corresponding standard deviation. The criterion error is mainly the total variation distance (TVD) which is calculated using the $L1$ distance by the Scheff\'e lemma (see for example \cite{Meister} page 129.).
\begin{eqnarray*}
\text{TVD}(p_{\phi},p_T) & = & \sup_{a<b}\left|dP_{\phi}([a,b]) - dP_T([a,b])\right| \\
  & =  & \frac{1}{2}\int{|p_{\phi}(x) - p_T(x)|dx}.
\end{eqnarray*}
We also provide for the Gaussian mixture the values of the (squared root of the) $\chi^2$ divergence between the estimated model and the true mixture, since it gave infinite values for the Weibull experiment. The $\chi^2$ criterion is defined by:
\[\chi^2(p_{\phi},p_T) = \int{\frac{\left(p_{\phi}(x)-p_T(x)\right)^2}{p_{\phi^*}(x)}dx}.\]
The use of a distance such as the $\chi^2$ divergence is due to its relative-error property. In other words, it calculates the error at a point relatively to its true value. Hence, errors at small values of the true density have their share in the overall error and are no longer negligible to points with higher density value. The total variation indicates the maximum error we might commit when calculating probabilities by replacing the true distribution by the estimated one. \\
We used different $\varphi-$divergences to estimate the parameters and compared the performances of the two methods of estimating a $\varphi-$divergence presented in this paper. For the Gaussian mixture, we used the Pearson's $\chi^2$ and the Hellinger divergences, whereas in the Weibull mixture, we used the Neymann's $\chi^2$ and the Hellinger divergences. For the MDPD, we used $a=0.5$; a choice which gave the best tradeoff between robustness and efficiency in the simulation results in \cite{Diaa}. We illustrate also the performance of the EM method in the light of our method, i.e. using initializations verifying conditions (\ref{eqn:TwoGaussMixCond}) for the Gaussian mixture and conditions (\ref{eqn:ConditionWeibullMix}) for the Weibull one. When outliers were added, these initializations did not always result in good results and the convergence of the proportion was towards the border $\eta=0.1$ or $1-\eta=0.9$. In such situations, the EM algorithm was initialized using another starting point manually. Last but not least, for the proximal term, we used $\psi(t)=\frac{1}{2}(\sqrt{t}-1)^2$.\\
We used the Nelder-Mead algorithm (see \cite{NelderMead}) for all optimization calculus. The method proved to be more efficient in our context than other optimization algorithms although it has a slow convergence speed. Such method is derivative-free and applies even if the the objective function is not differentiable which may be the case of the estimated divergence defined through (\ref{eqn:DivergenceDef}). The Nelder-Mead algorithm is known to give good results in problems with dimension at least 2 and does not perform well in dimension 1. We thus used Brent's method for the unidimensional optimizations. It is also a derivative-free method which works in a compact subset from $\mathbb{R}$ only. The calculus was done under the statistical tool \cite{Rtool}.\\
Numerical integrations were performed using the \texttt{distrExIntegrate} function of package \texttt{distrEx} in the Gaussian mixture. It is a slight modification of the standard \texttt{integrate} function in the R statistical tool which performs a Gauss-Legendre quadrature approximation whenever function \texttt{integrate} fails to converge. For the Weibull mixture, the previous function did not converge always, and function \texttt{integral} of package \texttt{pracma} was used. Although being very slow, it performs very well especially on unbounded integrations and "extremely bad-behavior" integrands.
\subsection{The two-component Gaussian mixture revisited}\label{Example:DivergenceMixture}
\begin{figure}[ht]
\centering
\includegraphics[scale = 0.3]{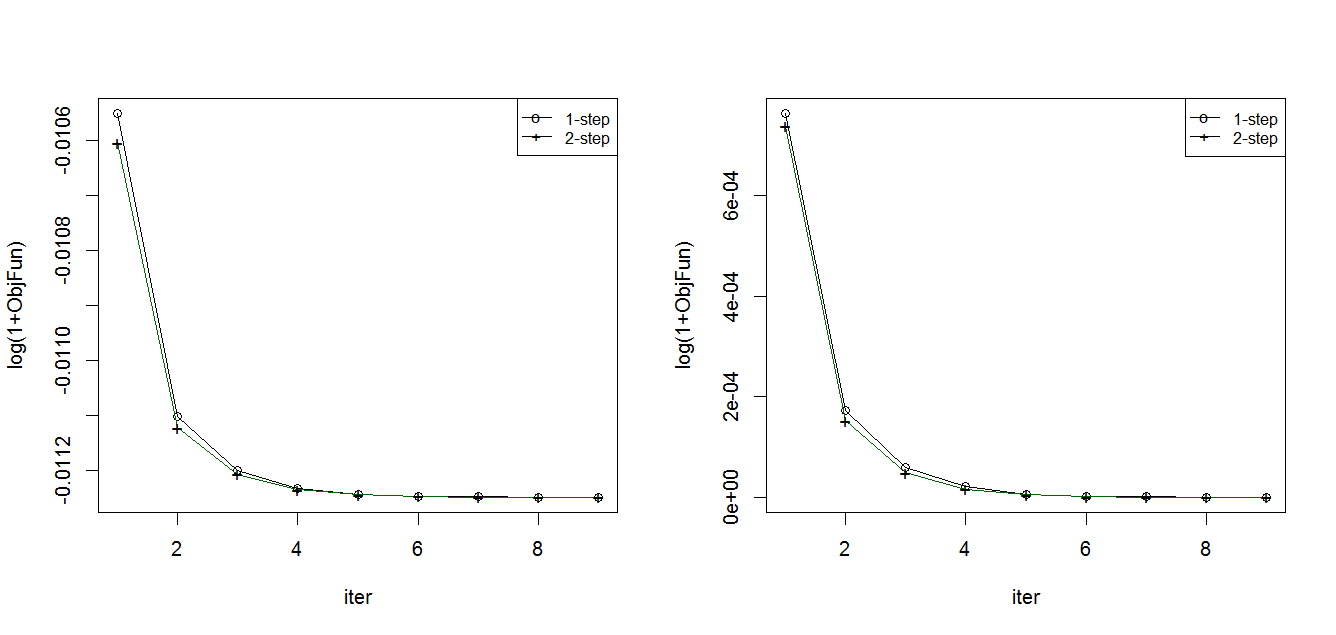}
\caption{Decrease of the (estimated) Hellinger divergence between the true density and the estimated model at each iteration in the Gaussian mixture. The figure to the left is the curve of the values of the kernel-based dual formula (\ref{eqn:NewDualForm}). The figure to the right is the curve of values of the classical dual formula (\ref{eqn:DivergenceDef}). Values are taken at a logarithmic scale $\log(1+x)$.}
\label{fig:DecreaseDivGaussChi2Chi2}
\end{figure}

We consider the Gaussian mixture (\ref{eqn:GaussMixModel}) presented earlier with true parameters $\lambda=0.35,\mu_1=2,\mu_2=1.5$ and fixed variances $\sigma_1=\sigma_2=1$. Since we are using a function error criterion, label-switching problems do not interfere. Figure (\ref{fig:DecreaseDivGaussChi2Chi2}) shows the values of the estimated divergence for both formulas (\ref{eqn:DivergenceDef}) and (\ref{eqn:NewDualForm}) on a logarithmic scale at each iteration of the algorithms (\ref{eqn:DivergenceAlgo}) and (\ref{eqn:DivergenceAlgoSimp1}, \ref{eqn:DivergenceAlgoSimp2}) until convergence. The 1-step algorithm refers to algorithm (\ref{eqn:DivergenceAlgo}), whereas 2-step refers to algorithm (\ref{eqn:DivergenceAlgoSimp1},\ref{eqn:DivergenceAlgoSimp2}). We omitted the initial point in order to produce a clear image of the decrease of the objective function. For the kernel-based dual formula, we used a Gaussian kernel with window calculate using Silverman's rule of thumb. Results are presented in table (\ref{tab:ErrGauss100RunEx}).\\
Contamination was done by adding in the original sample to the 5 lowest values random observations from the uniform distribution $\mathcal{U}[-5,-2]$. We also added to the 5 largest values random observations from the uniform distribution $\mathcal{U}[2,5]$. Results are presented in table (\ref{tab:ErrGauss00RunExOutl}).\\
It is clear that the both the MDPD and the kernel-based MD$\varphi$DE are more robust than the EM algorithm and the classical MD$\varphi$DE for both the Pearson's $\chi^2$ and the Hellinger divergences. Differences between the two choices of $\varphi-$divergences ($\chi^2$ and Hellinger) were not significant for the two $\varphi-$divergence-based estimators.

\begin{table}[!t]
\caption{The mean value of errors committed in a 100-run experiment with the standard deviation. No outliers are considered here. The divergence criterion is the Chi square divergence or the Hellinger. The proximal term is calculated with $\psi(t) = \frac{1}{2} (\sqrt{t}-1)^2$.}
\label{tab:ErrGauss100RunEx}
\centering
\begin{tabular}{|c|c|c|c|}
\hline
\multicolumn{2}{|c|}{\multirow{2}{2.5cm}{Estimation method}} & \multicolumn{2}{|c|}{Error criterion}\\
\cline{3-4}
 \multicolumn{2}{|c|}{} & $\sqrt{\chi^2}$ & TVD\\
 \hline
 \hline
\multicolumn{4}{|c|}{Chi square} \\
 \hline
 \hline
\multirow{2}{2.5cm}{Algorithm (\ref{eqn:DivergenceAlgo})}& MD$\varphi$DE & 0.108, sd = 0.052 & 0.061, sd = 0.029\\
& kernel-based MD$\varphi$DE & 0.118 , sd = 0.052 & 0.066 ,sd= 0.027\\
\hline
\multirow{2}{2.5cm}{Algorithm (\ref{eqn:DivergenceAlgoSimp1},\ref{eqn:DivergenceAlgoSimp2})}& MD$\varphi$DE & 0.108, sd = 0.052 & 0.061, sd = 0.029\\
& kernel-based MD$\varphi$DE & 0.118, sd = 0.051 & 0.066 ,sd= 0.027\\
\hline
\hline
\multicolumn{4}{|c|}{Hellinger} \\
\hline
\hline
\multirow{2}{2.5cm}{Algorithm (\ref{eqn:DivergenceAlgo})}& MD$\varphi$DE & 0.108, sd = 0.052 & 0.050 , sd=0.025\\
& kernel-based MD$\varphi$DE & 0.113, sd = 0.044 & 0.064 ,sd=0.025\\
\hline
\multirow{2}{3cm}{Algorithm (\ref{eqn:DivergenceAlgoSimp1},\ref{eqn:DivergenceAlgoSimp2})}& MD$\varphi$DE & 0.108, sd = 0.052 & 0.061, sd = 0.029\\
& kernel-based MD$\varphi$DE & 0.113, sd = 0.045 & 0.064 ,sd=0.025\\
\hline
\hline
\multicolumn{2}{|c|}{MDPD $a=0.5$ - Algorithm (\ref{eqn:DivergenceAlgo})} & 0.117, sd = 0.049 & 0.065, sd = 0.025 \\
\multicolumn{2}{|c|}{MDPD $a=0.5$ - Algorithm (\ref{eqn:DivergenceAlgoSimp1},\ref{eqn:DivergenceAlgoSimp2})}  & 0.117, sd = 0.047 & 0.065, sd = 0.025 \\
\hline
\hline
\multicolumn{2}{|c|}{EM} 	& 0.113, sd = 0.044 & 0.064 , sd = 0.025\\
\hline
\end{tabular}
\end{table}
\begin{table}[hp]
\caption{Error committed in estimating the parameters of a 2-component Gaussian mixture with $10\%$ outliers. The divergence criterion is the Chi square divergence or the Hellinger. The proximal term is calculated with $\psi(t) = \frac{1}{2} (\sqrt{t}-1)^2$.}
\label{tab:ErrGauss00RunExOutl}
\centering
\begin{tabular}{|c|c|c|c|}
\hline
\multicolumn{2}{|c|}{\multirow{2}{2.5cm}{Estimation method}} & \multicolumn{2}{|c|}{Error criterion}\\
\cline{3-4}
 \multicolumn{2}{|c|}{} & $\chi^2$ & TVD\\  \hline
 \hline
\multicolumn{4}{|c|}{Chi square} \\
 \hline
 \hline
\multirow{2}{2.5cm}{Algorithm (\ref{eqn:DivergenceAlgo})}& MD$\varphi$DE & 0.334, sd = 0.097 & 0.146,sd=0.036\\
 & kernel-based MD$\varphi$DE & 0.149 , sd = 0.059 & 0.084 ,sd=0.033\\
\hline
\multirow{2}{3cm}{Algorithm (\ref{eqn:DivergenceAlgoSimp1},\ref{eqn:DivergenceAlgoSimp2})}& MD$\varphi$DE & 0.333, sd = 0.097 & 0.149, sd = 0.033\\
& kernel-based MD$\varphi$DE & 0.149 , sd = 0.059 & 0.084, sd=0.033\\
\hline
\hline
\multicolumn{4}{|c|}{Hellinger} \\
\hline
\hline
\multirow{2}{2.5cm}{Algorithm (\ref{eqn:DivergenceAlgo})}& MD$\varphi$DE & 0.321, sd = 0.096 & 0.146, sd=0.034\\
& kernel-based MD$\varphi$DE & 0.155 , sd = 0.059 & 0.087 ,sd=0.033\\
\hline
\multirow{2}{3cm}{Algorithm (\ref{eqn:DivergenceAlgoSimp1},\ref{eqn:DivergenceAlgoSimp2})}& MD$\varphi$DE & 0.322, sd = 0.097 & 0.147, sd = 0.034\\
& kernel-based MD$\varphi$DE & 0.156 , sd = 0.059 & 0.087 ,sd=0.033\\
\hline
\hline
\multicolumn{2}{|c|}{MDPD $a=0.5$ - Algorithm (\ref{eqn:DivergenceAlgo})}  & 0.129, sd = 0.049 & 0.065, sd = 0.025 \\
\multicolumn{2}{|c|}{MDPD $a=0.5$ - Algorithm (\ref{eqn:DivergenceAlgoSimp1},\ref{eqn:DivergenceAlgoSimp2})}  & 0.138, sd = 0.053 & 0.078, sd = 0.030 \\
\hline
\hline
\multicolumn{2}{|c|}{EM} & 0.335, sd = 0.102 & 0.150, sd = 0.034 \\
\hline
\end{tabular}
\end{table}
\subsection{The two-component Weibull mixture model revisited}
We consider the Weibull mixture (\ref{eqn:WeibullMixture}) with $\phi_1 = 0.5, \phi_2 = 3$ and $\lambda = 0.35$ which are supposed to be unknown during the estimation procedure. We denote $\phi=(\phi_1,\phi_2)$ ($\alpha=(\alpha_1,\alpha_2)$, respectively) the shapes of the Weibull mixture model $p_{(\lambda,\phi)}$ ($p_{(\lambda,\alpha)}$, respectively). Contamination was done by replacing 10 observations of each sample chosen randomly by 10 i.i.d. observations drawn from a Weibull distribution with shape $\nu = 0.9$ and scale $\sigma = 3$. Results are presented in tables (\ref{tab:ErrWeibull100RunEx}) and (\ref{tab:ErrWeibull100RunOutliersEx}).\\ 
Manipulating the optimization procedure for the Neymann's $\chi^2$ was difficult because of the numerical integration calculus and the fact that for a subset of $\Phi$ (or $\Phi\times\Phi$ according to whether we use the estimator (\ref{eqn:DivergenceDef}) or the estimator (\ref{eqn:NewDualForm})) the integral term produces infinity, see paragraph \ref{subsec:WeibullMixEx} for more details. We therefore needed to keep the optimization from approaching the border in order to avoid numerical problems. For the Hellinger divergence, there is no particular remark.\\
For the case of the estimated divergence (\ref{eqn:DivergenceDef}), if $\gamma=-1$, i.e. the Neymann $\chi^2$, we need that $\alpha_1 < \phi_1/2$, otherwise the integral term is equal to infinity. In order to avoid numerical complications, we optimized over $\alpha_1\leq 0.05+\phi_1/2$. The value $0.05$ ensures a small deviation from the border. \\
For the case of the estimated divergence (\ref{eqn:NewDualForm}), we used a Gaussian kernel for the Hellinger divergence. For the Neymann's $\chi^2$ divergence, we used the Epanechnikov's kernel to avoid problems at infinity. Besides, it permits to integrate only over $[0,\max(Y)+w]$, where $w$ is the window of the kernel, instead of $[0,\infty)$. In order to avoid problems near zero, it is necessary that $\min(\phi_1,\phi_2)<1-\frac{1}{\gamma}=2$.\\
Experimental results show a clear robustness of the estimators calculated using the density power divergece (the MDPD) and the kernel-based MD$\varphi$DE in comparison to other estimators using the Hellinger divergence. When we are under the model, all estimation methods have the same performance. On the other hand, using the Neymann $\chi^2$ divergence, results are different in the presence of outliers. The classical MD$\varphi$DE calculated using formula (\ref{eqn:DivergenceDef}) shows better robustness than other estimators except for the MDPD, but is still not as good as the robustness of the kernel-based MD$\varphi$DE using the Hellinger or the MDPD.  Lack of robustness of the kernel-based MD$\varphi$DE is not very surprising since the influence function of the kernel-based MD$\varphi$DE is unbounded when we use the Neymann $\chi^2$ divergence in simple models such as the Gaussian model, see Example 2 in \cite{Diaa}.\\
In what concerns the proximal algorithm, there is no significant difference between the results obtained using the 1-step algorithm (\ref{eqn:DivergenceAlgo}) and the ones obtained using the 2-step algorithm (\ref{eqn:DivergenceAlgoSimp1},\ref{eqn:DivergenceAlgoSimp2}) using the Hellinger divergence. Differences appear when we used the Neymann $\chi^2$ divergence with the classical MD$\varphi$DE. This shows again the difficulty in handling the supermal form of the dual formal (\ref{eqn:DivergenceDef}).

\begin{table}[hp]
\caption{The mean value of errors committed in a 100-run experiment of a two-component Weibull mixture with the standard deviation. No outliers are considered. The divergence criterion is the Neymann's $\chi^2$ divergence or the Hellinger. The proximal term is taken with $\psi(t) = \frac{1}{2} (\sqrt{t}-1)^2$.}
\label{tab:ErrWeibull100RunEx}
\centering
\begin{tabular}{|c|c|c|}
\hline
\multicolumn{2}{|c|}{\multirow{2}{2.5cm}{Estimation method}} & Error criterion\\
\cline{3-3}
 \multicolumn{2}{|c|}{} & TVD\\
 \hline
 \hline
\multicolumn{3}{|c|}{Neymann Chi square} \\
 \hline
 \hline
\multirow{2}{2.5cm}{Algorithm (\ref{eqn:DivergenceAlgo})}& MD$\varphi$DE & 0.114 , sd = 0.032\\
									& kernel-based MD$\varphi$DE & 0.057, sd = 0.028\\
\hline
\multirow{2}{2.5cm}{Algorithm (\ref{eqn:DivergenceAlgoSimp1},\ref{eqn:DivergenceAlgoSimp2})}& MD$\varphi$DE &  0.131, sd =  0.042\\
									& kernel-based MD$\varphi$DE & 0.056, sd = 0.026\\
\hline
\hline
\multicolumn{3}{|c|}{Hellinger} \\
\hline
\hline
\multirow{2}{2.5cm}{Algorithm (\ref{eqn:DivergenceAlgo})}& MD$\varphi$DE & 0.059, sd = 0.024\\
									& kernel-based MD$\varphi$DE & 0.057, sd = 0.029 \\
\hline
\multirow{2}{2.5cm}{Algorithm (\ref{eqn:DivergenceAlgoSimp1},\ref{eqn:DivergenceAlgoSimp2})}& MD$\varphi$DE & 0.061, sd = 0.026\\
									& kernel-based MD$\varphi$DE & 0.057, sd = 0.029\\
\hline
\hline
\multicolumn{2}{|c|}{MDPD $a=0.5$ - Algorithm (\ref{eqn:DivergenceAlgo})}  & 0.056, sd = 0.029 \\
\multicolumn{2}{|c|}{MDPD $a=0.5$ - Algorithm (\ref{eqn:DivergenceAlgoSimp1},\ref{eqn:DivergenceAlgoSimp2})}  & 0.056, sd = 0.029 \\
\hline
\hline
\multicolumn{2}{|c|}{EM} & 0.059, sd = 0.024\\
\hline
\end{tabular}
\end{table}

\begin{table}[hp]
\caption{The mean value of errors committed in a 100-run experiment of a two-component Weibull mixture with the standard deviation. $10\%$ outliers are considered. The divergence criterion is the Neymann's $\chi^2$ divergence or the Hellinger. The proximal term is taken with $\psi(t) = \frac{1}{2} (\sqrt{t}-1)^2$.}
\label{tab:ErrWeibull100RunOutliersEx}
\centering
\begin{tabular}{|c|c|c|}
\hline
\multicolumn{2}{|c|}{\multirow{2}{2.5cm}{Estimation method}} & Error criterion\\
\cline{3-3}
 \multicolumn{2}{|c|}{} & TVD\\
 \hline
 \hline
\multicolumn{3}{|c|}{Neymann Chi square} \\
 \hline
 \hline
\multirow{2}{2.5cm}{Algorithm (\ref{eqn:DivergenceAlgo})}& MD$\varphi$DE & 0.085,  sd = 0.036\\
									& kernel-based MD$\varphi$DE & 0.138, sd = 0.066 \\
\hline
\multirow{2}{2.5cm}{Algorithm (\ref{eqn:DivergenceAlgoSimp1},\ref{eqn:DivergenceAlgoSimp2})}& MD$\varphi$DE & 0.096, sd = 0.057\\
									& kernel-based MD$\varphi$DE & 0.127, sd = 0.056\\
\hline
\hline
\multicolumn{3}{|c|}{Hellinger} \\
\hline
\hline
\multirow{2}{2.5cm}{Algorithm (\ref{eqn:DivergenceAlgo})}& MD$\varphi$DE & 0.120, sd = 0.034\\
									& kernel-based MD$\varphi$DE & 0.068, sd = 0.034 \\
\hline
\multirow{2}{2.5cm}{Algorithm (\ref{eqn:DivergenceAlgoSimp1},\ref{eqn:DivergenceAlgoSimp2})}& MD$\varphi$DE & 0.121, sd = 0.034\\
									& kernel-based MD$\varphi$DE & 0.068, sd = 0.034\\
\hline
\hline
\multicolumn{2}{|c|}{MDPD $a=0.5$ - Algorithm (\ref{eqn:DivergenceAlgo})}  & 0.060, sd = 0.029 \\
\multicolumn{2}{|c|}{MDPD $a=0.5$ - Algorithm (\ref{eqn:DivergenceAlgoSimp1},\ref{eqn:DivergenceAlgoSimp2})}  & 0.061, sd = 0.029 \\
\hline
\hline
\multicolumn{2}{|c|}{EM} & 0.129, sd = 0.046\\
\hline
\end{tabular}
\end{table}
\clearpage
\section{Conclusions}
We presented in this paper a proximal-point algorithm whose objective was the minimization of (an estimate of) a $\varphi-$divergence. The set of algorithms proposed here covers the EM algorithm. We provided in several examples a proof of convergence of the EM algorithm in the spirit of our approach. We also showed how we may prove convergence for the two estimates of the $\varphi-$divergence (\ref{eqn:DivergenceDef}) and (\ref{eqn:NewDualForm}) and for the density power divergence (\ref{eqn:DPD}). We reestablished similar results to the ones in \cite{Tseng} in the context of general divergences, and provided a new result by relaxing the identifiability condition on the proximal term. Our simulation results permit to conclude that the proximal algorithm works. The two-step algorithm (\ref{eqn:DivergenceAlgoSimp1},\ref{eqn:DivergenceAlgoSimp2}) showed in the most difficult situations considered here a slight deterioration in performance comparing to the original one (\ref{eqn:DivergenceAlgo}) which is very encouraging especially that the dimension of the optimization is reduced at each step. Simulations have shown again the robustness of $\varphi-$divergences and the density power divergence against outliers in comparison to the MLE. The algorithm could be used to calculate other divergence-based estimators such as \cite{BasuLindsay} and \cite{Beran} or R\'enyi pseudodistances (\cite{TomaAubin}). The role of the proximal term and its influence on the convergence of the algorithm were not discussed here and will be considered in a future work.

\bibliographystyle{IEEEtran}
\bibliography{IEEEabrv,Biblio}

\end{document}